\documentclass[sigconf]{acmart}

\usepackage{multirow}
\usepackage{array}
\usepackage{subcaption}
\usepackage{stfloats}

\newcommand{\revision}{\textcolor{black}}
\newcommand{\revisionb}{\textcolor{black}}

\AtBeginDocument{%
  \providecommand\BibTeX{{%
    \normalfont B\kern-0.5em{\scshape i\kern-0.25em b}\kern-0.8em\TeX}}}

\copyrightyear{2024}
\acmYear{2024}
\setcopyright{acmlicensed}\acmConference[CHI '24]{Proceedings of the CHI Conference on Human Factors in Computing Systems}{May 11--16, 2024}{Honolulu, HI, USA}
\acmBooktitle{Proceedings of the CHI Conference on Human Factors in Computing Systems (CHI '24), May 11--16, 2024, Honolulu, HI, USA}
\acmDOI{10.1145/3613904.3642146}
\acmISBN{979-8-4007-0330-0/24/05}

\acmConference[CHI '24]{Make sure to enter the correct
  conference title from your rights confirmation emai}
\acmPrice{15.00}
\acmISBN{978-1-4503-XXXX-X/18/06}

\AtBeginEnvironment{quote}{\small}
\begin{document}

\title{A Change of Scenery: Transformative Insights from Retrospective VR Embodied Perspective-Taking of Conflict With a Close Other}

\author{Seraphina Yong}
\affiliation{%
  \institution{University of Minnesota}
  \city{Minneapolis, MN}
  \country{USA}}
\email{yong0021@umn.edu}

\author{Leo Cui}
\affiliation{%
  \institution{University of Minnesota}
  \city{Minneapolis, MN}
  \country{USA}}
\email{cui00103@umn.edu}

\author{Evan Suma Rosenberg}
\affiliation{%
  \institution{University of Minnesota}
  \city{Minneapolis, MN}
  \country{USA}}
\email{suma@umn.edu}

\author{Svetlana Yarosh}
\affiliation{%
  \institution{University of Minnesota}
  \city{Minneapolis, MN}
  \country{USA}}
\email{lana@umn.edu}

\renewcommand{\shortauthors}{Yong, et al.}
\renewcommand{\shorttitle}{Transformative Insights from Retrospective VR Embodied Perspective-Taking of Conflict With a Close Other}

\begin{abstract}
Close relationships are irreplaceable social resources, yet prone to high-risk conflict. Building on findings from the fields of HCI, virtual reality, and behavioral therapy, we evaluate the unexplored potential of retrospective VR-embodied perspective-taking to fundamentally influence conflict resolution in close others. 
We develop a biographically-accurate Retrospective Embodied Perspective-Taking system (REPT) and conduct a mixed-methods evaluation of its influence on close others' reflection and communication, compared to video-based reflection methods currently used in therapy (treatment as usual, or TAU).
Our key findings provide evidence that REPT was able to significantly improve communication skills and positive sentiment of both partners during conflict, over TAU. 
The qualitative data also indicated that REPT surpassed basic perspective-taking by exclusively stimulating users to embody and reflect on both their own and their partner's experiences at the same level.
In light of these findings, we provide implications and an agenda for social embodiment in HCI design: conceptualizing the use of `embodied social cognition,' and envisioning socially-embodied experiences as an interactive context.
\end{abstract}




\maketitle

\section{Introduction}
Close relationships (e.g., family members, significant others, or close friends) are strong predictors of health and well-being \cite{pietromonaco_interpersonal_2017,deci_autonomy_2014}, but prone to conflict.
Conflict in close others has distinctly higher risk and impact resulting from the level of mutual dependence that is present in close relationships \cite{canary_relationship_1995}, but at the same time serves as a necessary obstacle for developing the deeper understanding of the other (also known as \textit{intersubjectivity} \cite{colvin_why_1997}) that is essential for maintaining close relationships. 
Compared to a task-based conflict where concrete situational aspects may be elevated over the personal, a conflict between close others may involve issues which intimately affect the relationship parties, such as management of personal resources or conflicting social preferences.

Existing technologies for conflict resolution, such as robot mediators and artificially-altered voice feedback, have been able to de-escalate conflicts and raise awareness of harmful actions by providing guidance on negative behavior \cite{shen_stop_2018,jung_using_2015} or decreasing feelings of anxiety during conflict \cite{costa_regulating_2018}. 
However, the depth of close relationships may necessitate tools that cut deeper into the process of socially-significant and longer-sustaining conflict.
Technology can support only to mid-conflict solutions, but also inter-conflict reflection. Current applications demonstrate technology's versatility in representing information for social reflection \cite{kim_dyadic_2020,saksono_social_2019,semsioglu_isles_2021,mcduff_affectaura_2012}.
In contrast to existing mid-conflict approaches, we posit that promoting deeper reflections in-between conflict occurrences may support the development of crucial transformative insights and approaches for handling nontrivial conflict in close relationships \cite{canary_relationship_1995}.

Perspective-taking skills are a central factor to conflict resolution in close relationships \cite{kellas_communicated_2017}. Conflicts involve a difference in perspective, and perspective-taking remedies this by improving problem-solving ability and understanding of partners' mental-emotional states \cite{rizkalla_roles_2008,nelson_perspective-taking_2017}.
Traditionally, conflict resolution therapy has used video-mediated retrospection to motivate reflection \cite{fichten_see_1984,wright_denial_1976}, but lacks evidence supporting its ability to stimulate perspective-taking or change conflict behavior in close others \cite{fichten_problem-solving_1983,fichten_videotape_1983,fichten_see_1984}.
Meanwhile, VR research has explored the use of embodied perspective-taking to induce empathy and prosocial behavior towards vulnerable populations \cite{yee_walk_2006,ingram_evaluation_2019,kishore_virtual_2021,ahn_effect_2013}, but this approach has so far been applied only to strangers.
The embodied perspective-taking process may look different when explored in close others, who have fundamentally different models of shared understanding \cite{colvin_why_1997} and distinct social biases \cite{nickerson_how_1999,kenny_accuracy_2001,fiedler_actor-observer_1995} that may affect baseline perceptions.
These parallels between the role of perspective-taking in close-other conflict resolution and the rise of social applications in VR-embodied perspective-taking inspire the current work.

Our work aims to understand how embodied perspective-taking that truly represents one's unique personal experience with a close other can impact reflection and communication during conflict between close others.
To achieve this goal, we introduce and evaluate a novel hybrid of embodied perspective-taking and video-based recall, \textit{retrospective embodied perspective-taking} (REPT), which enables one to take the perspective of a close other during a real past conversation. This is implemented by filming a 360\textdegree{} video from the user's partner's POV during the past conversation and enabling the user to re-experience that conversation from their partner's POV in immersive VR format.
We want to evaluate how REPT affects the context of close-other conflict resolution, compared to the media-based reflection method currently-used in conflict therapy: video-mediated recall which displays a split view of both partners in the conversation \cite{overall_attachment_2015}. Our research questions focus on social reflection, communication behavior, and comparison between the two forms of reflective media:
\begin{quote}
\textbf{RQ1:} How does REPT impact users' \textbf{reflections} on a conflict discussion with a close other and \textbf{perceptions of themselves and their partners} \textit{in comparison to} the traditional form of video-mediated recall?\\
\textbf{RQ2:} How does \textbf{subsequent communication} around the same conflict issue change as a result of using REPT, \textit{in comparison to} the traditional form of video-mediated recall?
\end{quote}
We investigated these questions by conducting a between-subjects, mixed-methods controlled study with 26 romantic partner dyads, divided into two conditions: REPT and the traditional split-view video \cite{overall_attachment_2015}, which we will refer to as treatment-as-usual, or TAU. 
Study participants engaged in an initial conflict discussion which was recorded to generate the reflective media, then completed a second discussion after reflecting on the generated media (REPT or TAU) through a structured interview. The specific contributions of this work are as follows:
\begin{itemize}
    \itemsep-0.1em 
    \item Empirical evidence that the VR format of retrospective embodied perspective-taking (REPT) significantly improves specific communication skills and positive sentiment during personally-meaningful conflict between close others, over traditional video-mediated reflection.
    \item Empirical demonstration that the REPT method can result in correction of strong attributional biases towards close others. Our results also show evidence for cognitive empathy, which previously found to be unaffected by VR experiences \cite{martingano_virtual_2021}.
    \item Implications and an agenda for new considerations of embodied experiences in HCI design: the use of \textbf{`embodied social cognition'} to facilitate user-generated social understanding, and envisioning \textbf{`embodied experience as interaction context'} in the design of future systems for social reflection.
\end{itemize}

\section{Background \& Related Work}

In the following section, we first detail the nature of close relationship conflict and its links to perspective-taking. We then follow with a survey of technology-mediated approaches to conflict resolution and perspective-taking, situating the motivation for the current work.

\subsection{Close Relationships and Conflict}
Close relationships fulfill basic psychological needs through their high levels of interdependence and relatedness \cite{canary_relationship_1995}, but are difficult to sustain due to inherent personal differences which produce conflict \cite{deci_autonomy_2014, canary_relationship_1995}.
Close others uniquely develop and share an \textit{intersubjective meaning context} that emerges from shared experiences, observation, and communication of each other's thoughts and feelings. This enhanced interpretive framework helps maintain the sense of relatedness \cite{colvin_why_1997}.

\subsubsection{Bias and Perspective-Taking}
However, this degree of closeness in close relationships also produces various conflict-inducing biases.
\textit{Overimputation}, which is the overestimation of the overlap between others' and one's own knowledge, occurs disproportionately in close relationships and is strong enough to offset the benefits of intersubjective accuracy and exacerbate differences between partners \cite{nickerson_how_1999,kenny_accuracy_2001}. 
The \textit{actor-observer attribution bias} and \textit{egocentric bias} also cause close others to more likely to blame themselves and each other instead of considering other factors in the situation \cite{thompson_judgments_1981,fiedler_actor-observer_1995,fiedler_language_1991}. 
Additionally, the perceived significance of a close relationship increases the weight of risk imposed by personal conflict issues, such as perceived incompatibilities \cite{canary_relationship_1995,kellas_communicated_2017}.

Perspective-taking, often defined as the ability to understand what another is thinking by putting oneself in their place, is a core component of reducing conflict by encouraging close others to acknowledge and confirm one another's perspectives during an interaction \cite{klimecki_role_2019,kellas_communicated_2017}.
In light of strongly-biased perceptions, this ability to reappraise the situation is particularly important. 
Perspective-taking also correlates with other key skills for conflict resolution, such as problem-solving, de-escalation, repairing emotions, and physiological attunement \cite{rizkalla_roles_2008, nelson_perspective-taking_2017}.
However, effective perspective-taking does not come easily; self-prompted perspective-taking in close others has shown backfiring effects by strengthening biases and reducing relationship satisfaction --- people who try to imagine a close other's perspective have been shown to overestimate the transparency of their own values and preferences to their partner \cite{vorauer_potential_2013}.


\subsubsection{Video-mediated Reflection in Therapy}
The use of video-based retrospection on previous conflict discussions has been used in conjunction with therapy to improve conflict resolution and close relationship maintenance \cite{fichten_videotape_1983,darling_seeing_2009,wright_denial_1976},
after the discovery that watching video replay of past conversations was able to reduce the \textit{actor-observer attribution bias} in strangers \cite{storms_videotape_1973,regan_empathy_1975}.
Even though this method continues to be used as a standard component of relationship therapy and research \cite{overall_attachment_2015},
studies investigating the effect of video-based reflection on relationship conflict have found no evidence of its effect on close partners' social evaluations and behavior \cite{fichten_see_1984,fichten_problem-solving_1983,fichten_videotape_1983}. 
\revision{Informed by work described in the following section, our proposed use of embodied perspective-taking aims to provide a more effective mechanism for close interpersonal conflict by integrating the links across parallel domains of research on perspective-taking.}

\subsection{Technology-driven Approaches to Conflict Resolution and Perspective-taking}
\label{subsec:tech}
Considering the importance of perspective-taking in conflict resolution, we survey the technology approaches that intersect with this domain.
HCI research has presented various tools for conflict management, and the VR field has studied embodied perspective-taking for promoting empathy in a multitude of contexts.
These areas have primarily advanced in parallel of one another, as yet without intersection.

\subsubsection{Conflict Resolution Approaches in HCI}
\label{subsubsec:conflict}
Work in the field of HCI has made a variety of conceptual and tangible contributions to address conflict management.
Investigation of how conflict occurs in a variety of online ecosystems, such as Wikipedia and online multiplayer gaming platforms, has generated design recommendations for these venues to resolve conflict, and ways to adapt offline conflict-resolution training to online contexts \cite{kittur_he_2007,billings_understanding_2010,jagannath_we_2020}.
Tools that directly assist with mediating interpersonal conflict, such as robot-mediated conflict, have also been explored. Robot interventions to repair negative behavior were able to increase awareness of harmful actions and reduce negative feelings in a team conflict \cite{jung_using_2015}, and robot facilitation of play and activity guidance improved constructive conflict resolution in preschool children \cite{shen_stop_2018}. Artificial alteration of one's own voice feedback into a calmer tone during a relationship conflict was also able to reduce feelings of anxiety and improve in-the-moment emotional regulation during the conflict \cite{costa_regulating_2018}.

These approaches to promote conflict resolution address the issue in diverse ways, ranging from clarifying the nature of the conflict, to supporting the training of conflict skills, to providing mechanisms that intervene in-the-moment to mitigate side effects of conflict and guide it. 
The current work embodies a blend of these methods: our approach is characterized by the design of a mechanism for inter-conflict use, meant to stimulate fundamental reflection and facilitate development of skills for future engagements in conflict issues.

\subsubsection{\revisionb{Technologies for Reflective Perspective-Taking}}
\label{subsubsec:pt_technologies}
Approaches in HCI for reflective perspective-taking have included Dyadic Mirror \cite{kim_dyadic_2020}, a wearable enabling parents to review interactions from their child's perspective and better understand their childrens' feelings; and ReliveInVR \cite{wang_again_2020}, a VR system allowing users to co-watch previous shared recreational VR experiences together from the same perspective, leading to improved understanding and social satisfaction.

Embodied perspective-taking in VR specifically has been applied in numerous social contexts, constrained mainly to ethical motivations. 
After introduction of the \textit{Proteus Effect} theory, which posits that people will behaviorally conform to the virtual representation of themselves based on avatar features (e.g., height, attractiveness) \cite{yee_proteus_2007}, multiple studies have followed which demonstrate increases of self-other merging, empathy, and prosocial helping behavior in participants who experienced VR-embodied perspective-taking of individuals belonging to various vulnerable demographics, ranging from individuals with colorblindness \cite{ahn_effect_2013}, disaster evacuees \cite{kors_curious_2020}, specific fictional characters \cite{bindman_am_2018,loon_virtual_2018}, the homeless \cite{herrera_building_2018}, and victims of violence \cite{ingram_evaluation_2019,kishore_virtual_2021,neyret_embodied_2020}.

\subsubsection{\revisionb{Elements of VR-Embodied Perspective-Taking}}
\label{subsubsec:embodiment}
\revisionb{Virtual (VR) embodiment employs mechanisms to support \textit{sense of embodiment} (SoE) in both \textit{other-oriented} perspective-taking and \textit{self-oriented} perspective-switching.
Slater's theory of SoE \cite{Slater2017} asserts that sense of embodiment is composed of three key components: \textit{self-location}, \textit{body ownership}, and \textit{agency}. These properties are implemented in our work (see \ref{subsubsec:implementation}) as well as many others to enhance virtual perspective-taking, having shown significant effects on aspects of social cognition such as reduction of implicit bias and increase in perceived affinity with outgroups \cite{MAISTER20156,bedder_mechanistic_2019}.
During \textit{VR other-embodiment} (embodying another person or character), matching the user's perspective in VR to the other body's first-person perspective facilitates self-location and body ownership by allowing the user to experience the environment through the target's body (e.g., head turning) \cite{slater_implicit_2017}. Adding more multisensory correlations (e.g., visuomotor) also enhances body ownership during perspective-taking of others' experiences, especially since the user may not have full control \cite{MAISTER20156}.
Agency is supported by allowing the user to affect the environment (e.g. by moving around), another common feature in VR other-embodiment work \cite{loon_virtual_2018, osimo_conversations_2015,ahn_effect_2013,bindman_am_2018,kishore_virtual_2021,neyret_embodied_2020}.
HCI work has also identified design features to support other-oriented empathy during embodied perspective-taking, such as incorporating real physical environmental elements, visceral bodily experiences, understanding of user role, and multisensory elements to affect emotion \cite{breathtaking2016,bindman_am_2018}.
In addition to \textit{other-oriented} understanding, virtual embodiment of an unfamiliar other is also used for \textit{self-oriented} perspective-switching during self reflection. In the context of self-counselling, people who embodied a different person in VR while talking to an avatar of themselves were able to detach from their habitual ways of thinking or self-criticism, which is a fundamental cognitive change \cite{osimo_conversations_2015,falconer_embodying_2016}.}
The involvement of implicit learning (e.g., reduction of implicit bias) \cite{slater_implicit_2017,MAISTER20156} and mirror neuron activation (sociophysiological synchrony) in VR perspective-taking \cite{august_fmri_2006,jackson_neural_2006} also support the role of subconscious processes in virtual embodiment.


\revisionb{Our current work on \textbf{close others} presents a distinct synthesis of the \textit{self-oriented} (self-counselling, reflection) and \textit{other-oriented} (empathy) perspective-taking contexts that have so far been explored separately in VR: a familiar relationship involves existing interrelated perspectives of \textit{both} the user and the target.
Our work is necessary to comprehend how VR embodiment and perspective-taking mechanisms extend to the close others context, as social perception in familiar others is known to \textit{differ} from strangers on dimensions core to embodied cognition in VR \cite{slater2016enhancing}: such as proxemics \cite{cristani2011}, brain activation patterns for empathy \cite{meyer_empathy_2012}, and metaperceptions of self \cite{boyes_metaperceptions_2007}.}

In synthesizing these multi-domain findings, we argue that the VR embodiment approach can be applied towards social perspective-taking in close relationships to instill meaningful changes in people's evaluations and conflict communication skills.
\revisionb{Research to-date has demonstrated both the context-dependent nature of conflict management (\ref{subsubsec:conflict}) and well-established potential for VR-embodied perspective-taking to instill fundamental changes in both self- and other-oriented reflection (\ref{subsubsec:embodiment}).
However, we do not yet understand how mechanisms of VR embodiment and perspective-taking work in the unique social context of familiar others, nor how meta-conflict approaches (e.g., reflection on conflicting perspectives) serve as a potential solution to close other conflict.
To build on these open areas, our work contributes a feasible implementation of authentic close-other perspective-taking and an empirical evaluation to identify social properties that can change and benefit from reflection through embodiment of close others.}



\begin{table*}[b]
\captionsetup{justification=centering}
\renewcommand{\arraystretch}{1}
\small{
  \centering
  \caption{
Demographic data of participating couples.
  }
\resizebox{\textwidth}{!}{%
\begin{tabular}{|ccccc|ccccc|}
\hline
\multicolumn{5}{|c|}{\textbf{REPT}}                                                                                                                                                                                                                                                                                                                                             & \multicolumn{5}{c|}{\textbf{TAU}}                                                                                                                                                                                                                                                                                                                                        \\ \hline
\multicolumn{1}{|c|}{\textbf{ID}} & \multicolumn{1}{c|}{\textbf{\begin{tabular}[c]{@{}c@{}}Age/Gender\\ (Partner A)\end{tabular}}} & \multicolumn{1}{c|}{\textbf{\begin{tabular}[c]{@{}c@{}}Age/Gender\\ (Partner B)\end{tabular}}} & \multicolumn{1}{c|}{\textbf{\begin{tabular}[c]{@{}c@{}}Orientation/\\ Duration (mos)\end{tabular}}} & \textbf{Conflict topic}                                                                 & \multicolumn{1}{c|}{\textbf{ID}} & \multicolumn{1}{c|}{\textbf{\begin{tabular}[c]{@{}c@{}}Age/Gender\\ (Partner A)\end{tabular}}} & \multicolumn{1}{c|}{\textbf{\begin{tabular}[c]{@{}c@{}}Age/Gender\\ (Partner B)\end{tabular}}} & \multicolumn{1}{c|}{\textbf{\begin{tabular}[c]{@{}c@{}}Orientation/\\ Duration (mos)\end{tabular}}} & \textbf{Conflict topic}                                                               \\ \hline
\multicolumn{1}{|c|}{V01}         & \multicolumn{1}{c|}{27 M}                                                                         & \multicolumn{1}{c|}{24 F}                                                                         & \multicolumn{1}{c|}{Heterosexual/24}         & \begin{tabular}[c]{@{}c@{}}Money \\ management\end{tabular}                             & \multicolumn{1}{c|}{T01}         & \multicolumn{1}{c|}{20 F}                                                                         & \multicolumn{1}{c|}{21 M}                                                                         & \multicolumn{1}{c|}{Heterosexual/34}         & Value conflict                                                                        \\
\multicolumn{1}{|c|}{V02}         & \multicolumn{1}{c|}{22 F}                                                                         & \multicolumn{1}{c|}{25 M}                                                                         & \multicolumn{1}{c|}{Heterosexual/32}         & \begin{tabular}[c]{@{}c@{}}Time \\ management\end{tabular}                              & \multicolumn{1}{c|}{T02}         & \multicolumn{1}{c|}{20 F}                                                                         & \multicolumn{1}{c|}{20 M}                                                                         & \multicolumn{1}{c|}{Heterosexual/18}         & \begin{tabular}[c]{@{}c@{}}Individual\\ problem affecting\\ relationship \end{tabular} \\[15pt]
\multicolumn{1}{|c|}{V03}         & \multicolumn{1}{c|}{36 F}                                                                         & \multicolumn{1}{c|}{36 M}                                                                         & \multicolumn{1}{c|}{Heterosexual/156}         & \begin{tabular}[c]{@{}c@{}}Communication \\ issues\end{tabular}                         & \multicolumn{1}{c|}{T03}         & \multicolumn{1}{c|}{20 F}                                                                         & \multicolumn{1}{c|}{20 F}                                                                         & \multicolumn{1}{c|}{Same-sex/24}             & \begin{tabular}[c]{@{}c@{}}Communication\\ issues\end{tabular}                        \\[10pt]
\multicolumn{1}{|c|}{V04}         & \multicolumn{1}{c|}{19 F}                                                                         & \multicolumn{1}{c|}{22 M}                                                                         & \multicolumn{1}{c|}{Heterosexual/8}         & Jealousy                                                                                & \multicolumn{1}{c|}{T04}         & \multicolumn{1}{c|}{27 M}                                                                         & \multicolumn{1}{c|}{26 F}                                                                         & \multicolumn{1}{c|}{Heterosexual/96}         & \begin{tabular}[c]{@{}c@{}}Money\\ management\end{tabular}                            \\
\multicolumn{1}{|c|}{V05}         & \multicolumn{1}{c|}{21 M}                                                                         & \multicolumn{1}{c|}{20 F}                                                                         & \multicolumn{1}{c|}{Heterosexual/33}         & \begin{tabular}[c]{@{}c@{}}Social \\ relationships\end{tabular}                         & \multicolumn{1}{c|}{T05}         & \multicolumn{1}{c|}{20 NB}                                                                        & \multicolumn{1}{c|}{19 F}                                                                         & \multicolumn{1}{c|}{Non-binary/10}           & Sex                                                                                   \\[10pt]
\multicolumn{1}{|c|}{V06}         & \multicolumn{1}{c|}{20 F}                                                                         & \multicolumn{1}{c|}{20 F}                                                                         & \multicolumn{1}{c|}{Same-sex/7}             & \begin{tabular}[c]{@{}c@{}}Social \\ relationships\end{tabular}                         & \multicolumn{1}{c|}{T06}         & \multicolumn{1}{c|}{26 M}                                                                         & \multicolumn{1}{c|}{26 F}                                                                         & \multicolumn{1}{c|}{Heterosexual/72}         & \begin{tabular}[c]{@{}c@{}}Money \\ management\end{tabular}                           \\[10pt]
\multicolumn{1}{|c|}{V07}         & \multicolumn{1}{c|}{20 F}                                                                         & \multicolumn{1}{c|}{20 M}                                                                         & \multicolumn{1}{c|}{Heterosexual/5}         & Jealousy                                                                                & \multicolumn{1}{c|}{T07}         & \multicolumn{1}{c|}{20 F}                                                                         & \multicolumn{1}{c|}{21 M}                                                                         & \multicolumn{1}{c|}{Heterosexual/10}         & Decision-making                                                                       \\[6pt]
\multicolumn{1}{|c|}{V08}         & \multicolumn{1}{c|}{26 F}                                                                         & \multicolumn{1}{c|}{25 NB}                                                                        & \multicolumn{1}{c|}{Non-binary/49}           & \begin{tabular}[c]{@{}c@{}}Demonstration\\ of affection\end{tabular}                    & \multicolumn{1}{c|}{T08}         & \multicolumn{1}{c|}{20 M}                                                                         & \multicolumn{1}{c|}{18 F}                                                                         & \multicolumn{1}{c|}{Heterosexual/10}         & \begin{tabular}[c]{@{}c@{}}Time \\ management\end{tabular}                            \\[10pt]
\multicolumn{1}{|c|}{V09}         & \multicolumn{1}{c|}{21 M}                                                                         & \multicolumn{1}{c|}{21 F}                                                                         & \multicolumn{1}{c|}{Heterosexual/30}         & \begin{tabular}[c]{@{}c@{}}Time \\ management\end{tabular}                              & \multicolumn{1}{c|}{T09}         & \multicolumn{1}{c|}{20 M}                                                                         & \multicolumn{1}{c|}{19 F}                                                                         & \multicolumn{1}{c|}{Heterosexual/10}         & \begin{tabular}[c]{@{}c@{}}Family \\ relationships\end{tabular}                       \\[10pt]
\multicolumn{1}{|c|}{V10}         & \multicolumn{1}{c|}{29 M}                                                                         & \multicolumn{1}{c|}{24 F}                                                                         & \multicolumn{1}{c|}{Heterosexual/3}         & Sex                                                                                     & \multicolumn{1}{c|}{T10}         & \multicolumn{1}{c|}{31 F}                                                                         & \multicolumn{1}{c|}{30 M}                                                                         & \multicolumn{1}{c|}{Heterosexual/36}         & Potential children                                                                    \\[5pt]
\multicolumn{1}{|c|}{V11}         & \multicolumn{1}{c|}{22 F}                                                                         & \multicolumn{1}{c|}{24 M}                                                                         & \multicolumn{1}{c|}{Heterosexual/3}         & Commitment                                                                              & \multicolumn{1}{c|}{T11}         & \multicolumn{1}{c|}{21 F}                                                                         & \multicolumn{1}{c|}{21 M}                                                                         & \multicolumn{1}{c|}{Heterosexual/4}         & \begin{tabular}[c]{@{}c@{}}Time \\ management\end{tabular}                            \\[10pt]
\multicolumn{1}{|c|}{V12}         & \multicolumn{1}{c|}{20 F}                                                                         & \multicolumn{1}{c|}{20 F}                                                                         & \multicolumn{1}{c|}{Same-sex/42}             & \begin{tabular}[c]{@{}c@{}}Time \\ management\end{tabular}                              & \multicolumn{1}{c|}{T12}         & \multicolumn{1}{c|}{21 M}                                                                         & \multicolumn{1}{c|}{21 F}                                                                         & \multicolumn{1}{c|}{Heterosexual/36}         & \begin{tabular}[c]{@{}c@{}}Communication \\ issues\end{tabular}                       \\[10pt]
\multicolumn{1}{|c|}{V13}         & \multicolumn{1}{c|}{28 M}                                                                         & \multicolumn{1}{c|}{37 F}                                                                         & \multicolumn{1}{c|}{Heterosexual/54}         & \begin{tabular}[c]{@{}c@{}}Individual \\ problem affecting \\ relationship\end{tabular} & \multicolumn{1}{c|}{T13}         & \multicolumn{1}{c|}{28 F}                                                                         & \multicolumn{1}{c|}{34 M}                                                                         & \multicolumn{1}{c|}{Heterosexual/24}         & \begin{tabular}[c]{@{}c@{}}Household \\ management\end{tabular}                       \\ \hline
\end{tabular}%
}
\label{table:participant}
}
\end{table*}

\section{Methods}
We conducted a between-subjects experiment consisting of two in-person sessions with participant pairs. The two conditions evaluated were VR-based \textit{REPT}, and \textit{TAU} (treatment as usual; traditional video replay of the conflict).
\revision{The objective of our formative investigation is to evaluate how social reflection and conflict behavior are affected by embodied perspective-taking (requiring the use of immersive VR \cite{yee_walk_2006}) in comparison to the standard reflection format currently used in conflict resolution therapy, which is video with a split-view of both partners \cite{overall_attachment_2015} (seen in \autoref{fig:video}c). Our choice to evaluate immersive embodiment in particular comes from strong evidence in its ability to facilitate perspective-taking and more fundamental changes in thinking about personal problems \cite{osimo_conversations_2015,yee_walk_2006}.}

In the first session, pairs of close others engaged in an in-person initial conflict discussion recorded for use in one of the two randomly-assigned conditions.
In the second session, participants returned to re-experience the initial conflict discussion through one of the two conditions, engage a structured interview, and engage in a second conflict discussion.

\begin{figure*}[b]
  \centering
  \includegraphics[width=\linewidth]{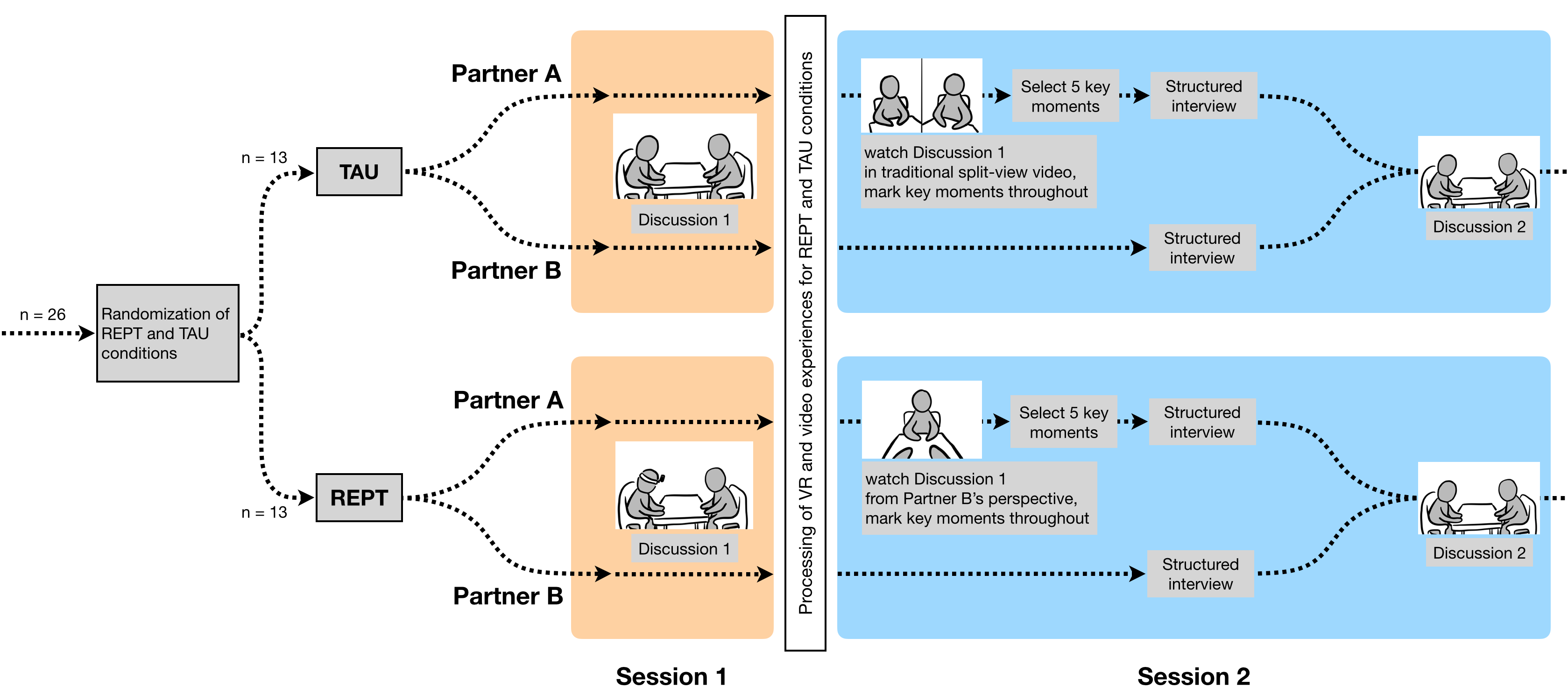}
  \caption{Flow diagram of study procedure including distribution of conditions, sessions, and task order between participants (both members of a couple). Discussions 1 and 2 occurred between both partners, while the other tasks were completed separately. See \autoref{session1} for details on Partner A and B.}
  \label{fig:procedure}
  \Description{Flow diagram of study procedure including distribution of sessions and tasks between participants (both members of a couple).}
\end{figure*}

\subsection{Participants}
We recruited 26 pairs of close others who (1) identified as in a committed romantic relationship and (2) specified a significant relationship-relevant conflict topic they were willing to discuss (\autoref{table:participant}).
We recruited only participants who expressed sincere intent to discuss after acknowledging the gravity of the chosen conflict (see \ref{subsec:ethics} for details).
\revision{Our recruitment methods included physical flyers with digital sign-up links distributed locally throughout a large city, and advertisement at a major university through cross-departmental mailing lists and research experience program.} 
We recruited close others as couples due the level of involvement and commitment between partners in a romantic relationship being a confident indicator of a close relationship.
Couples with diverse gender orientations were included in this work due to the lack of inherent association between gender and close relationships \cite{winstead_gender_1997}, which is the dynamic of focus.
Our sample included 21 heterosexual couples, four same-sex couples, and one non-binary couple.
Participants were between 18 and 36 years of age (M = 23.44, SD = 4.81);
 relationship duration ranged from 3 months to 13 years.
\revisionb{13 pairs participated in each condition; see Appendix \ref{subsec:power} for details on sample size. ANCOVA tests performed to assess potential interactions from age, gender, and relationship duration on our results (\autoref{sec:results}) found no significant effects.}
Participants were informed that the goal of the study was to understand how couples reflect on their communication through different types of technology.
All participants were compensated \$40.00 for their participation across the two in-person sessions, which totaled approximately 2.5 hours 
\revision{(20-30 minutes for Session 1, $\sim$1.5-2 hours for Session 2).}

\subsection{Procedure \& Implementation}
\label{subsec:procedure}
The study procedure spanned three phases: the first in-person session (Session 1) in which an initial conflict discussion was recorded, between-session processing of the retrospection media, and the second in-person session (Session 2) which includes the retrospection/re-experiencing of the first conflict discussion, interview, and second conflict discussion.

Both of the conflict discussion procedures took the same format and were adapted from the Markman-Cox procedure \cite{cox1991marital}, where after arriving in-person, couples were given $\leq$10 minutes to first privately confirm the topic of conflict they would like to discuss in the study. They were then instructed to discuss the topic with the goal of trying to understand and resolve the conflict, and left alone in the room for a 10-15 minute discussion which was video-recorded with fixed camera. In couples assigned to the REPT condition, the discussion was also recorded in 360\textdegree\ video from a head-mounted camera on one of the participants (see \autoref{fig:video}a).

\subsubsection{Session 1: Initial Conflict Discussion \& Recording}\label{session1}

In the first session of the study, both members of the couple arrived at the lab, read, and signed consent forms in a private room. \revisionb{They were then instructed to confirm the topic of conflict to discuss (which remains the same through both discussions in the study) and left alone in the room to allow for private discussion.} When finished, participants wrote the discussion topic they selected together on a sheet of paper and opened the door of the room to indicate they were ready to move on.

The experimenter then explained the conflict discussion phase of the study to participants. They were to discuss their chosen conflict topic for 10 minutes, with the goal of relating the topic to their relationship and attempting to resolve it. Participants would be left alone for the conversation with a timer in the room to let them to know how much time had passed. They were informed that if at 10 minutes they were still in the middle of talking, they could take the time to wrap up the conversation and notify the experimenter once they were finished.

\begin{figure*}[t]
  \centering
  \includegraphics[width=\linewidth]{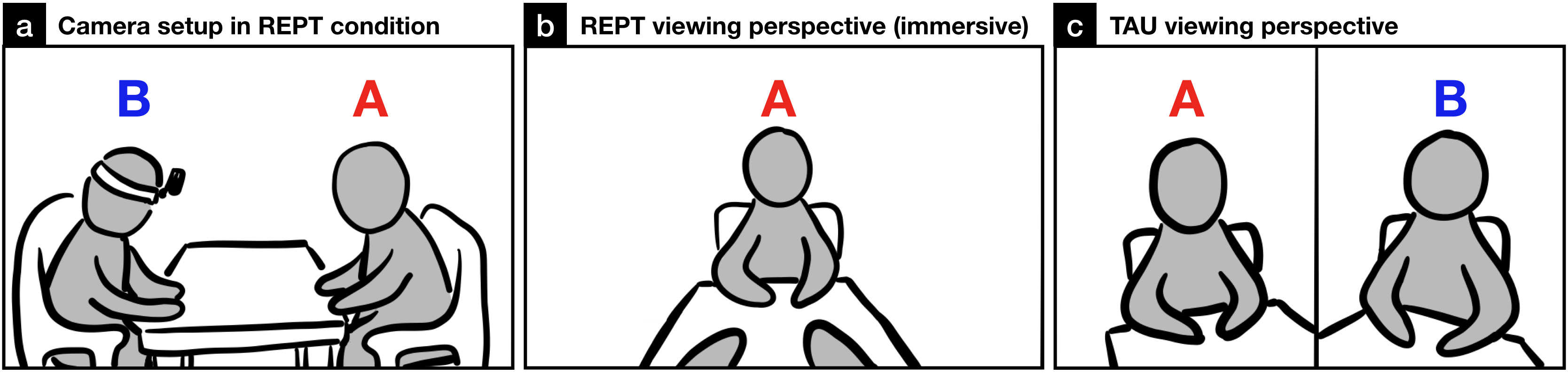}
  \caption{REPT (VR) and TAU (split-view video) recording setup and resulting views. \textbf{(a)} REPT condition recording setup with Partner B wearing the head-mounted camera. The TAU condition setup was exactly the same but without the head-mounted camera. \textbf{(b)} The REPT condition view extracted from the 360\textdegree\ recording, seen during Partner A's perspective-taking of Partner B. \textbf{(c)} The TAU condition view, with both partners visible.}
  \label{fig:video}
  \Description{REPT (VR) and TAU (traditional video) recording setup and resulting views. \textbf{(a)} REPT condition physical setup with Partner B wearing the head-mounted camera. The TAU condition setup was exactly the same but without the head-mounted camera. \textbf{(b)} The REPT condition view extracted from the 360\textdegree\ recording, seen during Partner A's perspective-taking of Partner B. \textbf{(c)} The TAU condition view, with both partners visible.}
\end{figure*}

Fixed-camera recording was set up after the topic selection and before the conflict discussion. Participants sat across from each other, and discussion recordings captured full-frontal views of each partner (\autoref{fig:video}c). In the REPT condition, one randomly-selected partner was fitted with a head-mounted GoPro Max 360 camera positioned and angled to the partner's eye-level, and the conversation was recorded in 360\textdegree\ video from that partner's perspective (\autoref{fig:video}a).
We will refer to this partner as \textbf{\textit{Partner B}} to explain the rest of the procedure.
Wearing the camera did not have any observable effect on the seriousness or depth of conversation between partners in our study. The other partner, \textbf{\textit{Partner A}}, is the primary participant to re-experience the discussion in Session 2, via one of the two conditions (see \autoref{fig:procedure}).
\revisionb{Because this is the first exploration of REPT, \textbf{only one person} from each couple experienced the intervention to initially observe REPT's direct effects on a single user.
Had both partners experienced the intervention, the possibility of synergistic effects during partner interaction in the post-discussion would conflate direct effects of REPT on an individual user with secondary effects from a REPT-influenced partner and prevent direct evaluation.}

\subsubsection{Implementations of Study Conditions}
\label{subsubsec:implementation}
Below, we will describe how the recordings of the Session 1 conflict discussion were processed for each condition. The resulting forms of media (VR or traditional video) were experienced by one of the partners in each couple during Session 2.

For the \textbf{REPT condition} which involves VR-based perspective-taking via 360\textdegree\ video, we extracted 360\textdegree\ footage and 6DoF ambisonic (spatial) audio from the GoPro Max 360 recording and processed them into a VR experience from the perspective of \textit{Partner B} using the Unity game engine. The resolution of the 360\textdegree\ video was 5.6K (5376x2688) at 30fps to maintain visual realism when accommodating the expanded field of view in VR. The ambisonic spatial audio was integrated in the egocentric perspective of \textit{Partner B}. We used a HTC Vive Pro headset for VR, which has a resolution of 2880x1600 and 120Hz refresh rate. Because the video was filmed from \textit{Partner B's} perspective, the other partner (\textit{Partner A}) would re-experience this discussion in Session 2 to enact perspective-taking of a close other (see \autoref{fig:video}b).

\revisionb{The design of the REPT system accounts for SoE mechanisms from prior work (see \ref{subsubsec:embodiment}), close others' increased sensitivity to physical cues \cite{sternglanz_reading_2004,visconti_di_oleggio_castello_shared_2021} and reproducibility of application. The dynamic perspective of \textit{Partner B} is superimposed onto \textit{Partner A} to facilitate self-location and body ownership.
The ensuing visuomotor illusions and use of readily-available 6DoF audio (human sounds) add multisensory correlations to enhance visceral embodiment \cite{breathtaking2016}, while \textit{Partner A} retains agency to visually navigate \textit{Partner B}'s perceived environment by moving their head.
We chose 360\textdegree\ video over other options (e.g., volumetric video, virtual body, real-time motion capture) as the best option for a reproducible implementation of close-other perspective-taking.
Close others have heightened sensitivity to nonverbal cues, constituting small changes in physical behavior \cite{sternglanz_reading_2004}, and neural-level sensitivity to familiar faces \cite{visconti_di_oleggio_castello_shared_2021}. Currently-accessible volumetric video solutions do not retain enough visual fidelity to discern social microgestures, and face familiarity increases the uncanniness of virtual faces \cite{diel_familiarity_2022}. We prioritized preserving perception of \textit{Partner B}'s physical behavior, excluding real-time motion capture which cannot reconcile the virtual user's actions with another body's \cite{neyret_embodied_2020}.}

For the \textbf{TAU condition}, footage of each partner in the discussion was recorded with the fixed cameras at 1080p, synchronized, and edited to split the view vertically, showing both partners (see \autoref{fig:video}c). In this condition, the designated re-experiencer (\textit{Partner A}) would watch this video in Session 2 instead of the REPT VR experience.

\subsubsection{Session 2: Retrospection, Interview, \& Post-Intervention Discussion}
Session 2 was completed between two days to one week after the first session to allow for the above implementation of the study conditions.
After participants returned for Session 2, each member of the couple was led to a separate room. \textit{Partner A} (defined in above sections) re-experienced the Session 1 discussion either in VR from the perspective of their partner (REPT), or in the TAU view showing both partner's perspectives on a computer screen.
To trigger reflection and measure relevant components of social evaluation (RQ1), we adapted Schulz and Waldinger's video recall procedure \cite{schulz2004looking} for \textit{Partner A}'s rewatching of the discussion (See \autoref{measures} for details on the adaptation of this measure).

\textbf{\textit{Retrospection.}} Prior to watching the discussion, participants in both conditions were given a handheld button and instructions to click the button whenever they felt they were experiencing significant thoughts or feelings during the watching experience.
The button would mark the time in the video at which it was clicked, but did not provide any sensory feedback to intrude on the participant's experience.
After the participant (\textit{Partner A}) re-experienced the entire discussion either through their partner's perspective (REPT) or via traditional video (TAU), they reviewed the moments they had marked during the viewing and selected the five they perceived to be the most significant and educational.

\textbf{\textit{Structured Interview.}} The proceeding interview with both partners was structured around these five moments selected by \textit{Partner A}. 
This was done in separate rooms for both \textit{Partner A and B}, by different experimenters (see \autoref{fig:procedure} for conceptual model).
The interview procedure builds on the widely used social interaction paradigm engineered by Ickes and colleagues \cite{ickes_naturalistic_1986}, which asks participants to indicate what they believe their partners were thinking and feeling during moments of their own videotaped interaction.
We adapt Schulz and Waldinger's procedure, which builds directly on Ickes' paradigm \cite{schulz2004looking}, for selecting the five significant moments and collecting participants' affect ratings.

The interview was iterative and the same following procedure was repeated for each of the five moments selected by \textit{Partner A}. 
A 20 second video clip encompassing the 10 seconds preceding and following the selected moment was played for the participant, then they rated their own and their partner's emotional experience during that part of the discussion separately on a scale from -5 to +5 (negative to positive affect) \cite{schulz2004looking}.
Afterwards, participants were asked to describe those thoughts and feelings during that moment for their partner and themselves.
This interview process was repeated for each of the five moments selected by \textit{Partner A}.

Both \textit{Partner A and B} in the couple experienced the above interview, but only \textit{Partner A} was asked the following additional questions: (1) \textbf{why they selected the moment} while re-experiencing the discussion, (2) whether there were \textbf{changes in evaluations} of their own/partners' thoughts and feelings a result of re-experiencing the discussion, and (3) whether they had \textbf{new goals or strategies} for how to approach future conversations with their partner.

\textbf{\textit{Post-Intervention Discussion.}} After the interview, participants were reunited in the same room as Session 1, and engaged in a second conflict discussion on the same topic following the format of Discussion 1. They received compensation at the end of the study.

\subsection{Measures}\label{measures}
Participants responded to questions about their reflections of the conflict discussion and perceptions of themselves and their partners (\textbf{RQ1}).
As conflict discussions occurred before and after the re-experiencing and reflection process, we also analyzed pre/post effects of the interventions on conflict behavior (\textbf{RQ2}).
We included the following measures:

\revisionb{\textit{Qualitative User Reflections.}
To develop in-depth understandings of users' reflections on their interactions, themselves, and their partners (\textbf{RQ1}), we conducted structured interviews with both participants (see \autoref{subsec:procedure}) where
each partner described in detail what they thought themselves and their partners were thinking and feeling during each of the five discussion moments that was played back to them. 
\textit{Partner A}'s portion included additional evaluations of the discussion and their relationship since they had selected the moments which were used in the interview. We performed thematic analysis on all 26 interviews with \textit{Partner A} of each couple, starting from open codes which were integrated into themes representing the types of reflections participants from each condition revealed through their dialogue.}

\textit{Empathic Accuracy Rating.}
We adapted Schulz and Waldinger's procedure for measuring empathic accuracy through quantitative affect ratings \cite{schulz2004looking}, which involves partners selecting high-affect moments during a conflict discussion, then both partners rating each other's emotional experience during each of those moments from -5 (very negative emotion) to +5 (very positive emotion) on an 11-point scale with discrete values and zero (neutral) in the center.
\revisionb{This is used as a quantitative component of \textbf{RQ1}, in addition to our in-depth structured interviews.}

Our adaptation involved \textit{Partner A} selecting five significant moments while re-experiencing the first conflict discussion, then both partners rating each other's emotional experience at each moment using the above scale. This is considered an affective dimension of \textit{empathic accuracy} \cite{schulz2004looking,ickes_naturalistic_1986}, defined as how accurately one can infer another's feelings.

\begin{table*}[hbt]
  \caption{Interactional Dimensions Coding System (IDCS) -- Definitions for each dimension.}
  \label{tab:idcs}
  \small
  \renewcommand{\arraystretch}{1.3}
  \begin{tabular}{>{\raggedright}p{0.3\linewidth}>{\raggedright\arraybackslash}p{0.65\linewidth}}
    \toprule
    \textbf{Dimension}&\textbf{Definition}\\
    \midrule
    Positive Affect&Positivity of facial expressions, body positioning, and emotional tone of voice\\
    Negative Affect&Negativity of facial expressions, body positioning, and emotional tone of voice\\
    Problem-solving Skills&Ability to define problem and attempt to work toward a mutually satisfactory solution\\
    Denial&Active rejection of problem existence or personal responsibility\\
    Dominance&Control exerted by individual over partner, such as forceful, monopolizing, or coercive behavior\\
    Support/Validation&Positive listening and speaking skills that demonstrate support and understanding to partner\\
    Conflict&Features of expressed struggle between partners, including tension, hostility, disagreement, antagonism\\
    Withdrawal&Avoidance of interaction or of problem discussion, such as evasion, retreat, or unresponsiveness\\
    Communication Skills&Ability to listen to partner and communicate thoughts and feelings constructively\\
    Positive Escalation \textit{(Dyadic)}&Chain reactions, snowballing, or triggering of positive behaviors between partners\\
    Negative Escalation \textit{(Dyadic)}&Chain reactions, snowballing, or triggering of negative behaviors between partners\\
  \bottomrule
\end{tabular}
\end{table*}
\textit{Interactional Dimensions Coding System.}
\revisionb{We used the Interactional Dimensions Coding System (IDCS) \cite{julien1987interactional} to analyze cross-condition effects of the interventions on communication behavior (\textbf{RQ2}) through observational ratings of pre/post conflict discussion footage.}
IDCS is a widely-used coding system to assess both affective (negative/positive emotion) and content-related (e.g., problem-solving skills, support, communication skills) dimensions of interactions between close others \cite{kerig_couple_2004}. 
We used all nine \textit{individual dimensions} and two \textit{dyadic dimensions} (See \autoref{tab:idcs}) from IDCS to assess each partner's verbal and nonverbal behaviors in both discussions.
Individual dimensions are applied separately to each partner, while dyadic dimensions are measured for each pair.
Each dimension was coded on a scale from 1 to 9 based on frequency and intensity of adherence to content criteria in the IDCS training manual, which contains detailed descriptions and examples of each dimension.
Two of the authors coded the video data based on the manual, with highly satisfactory inter-rater reliabilities (Pearson's r) ranging from $r=0.81\sim\ 0.96$ across all dimensions.

\subsection{Ethical Considerations}
\label{subsec:ethics}
Due to the involvement of interpersonal conflict in this work, we made several considerations during the design of this study to minimize harm to participants (in addition to review and approval of the study by the authors' institutional IRB).
To ensure that participants were invested in addressing a significant topic of conflict and aware of the implications of discussing a conflict, we employed a pre-study screening questionnaire which detailed the conflict discussion required in the study and required individuals to list and describe in a few short sentences at least one conflict they were interested in addressing, as well as confirm their intent to have a genuine discussion. We did not recruit participants who were not able to provide these descriptions, and required both partners' consent to participation for successful recruitment.

During the conflict discussions, we arranged the physical environment to resemble a living space and designed the procedure so that participants were alone during discussion. This was intended to preserve participants' comfort and sense of privacy as much as possible during the conflict conversations, which often required vulnerability.
The instructions for the conflict discussion reminded them that the objective was to understand each others' differences and attempt to address the conflict; we did not require or encourage negative expressions of conflict. Finally, we actively debriefed all participants, giving them the opportunity to process the experience.

Our findings show deeper understandings between partners and changes in communication. We intend for this knowledge to support relationships where members benefit from sustainable engagement; applications of our findings should carefully consider how different relationship contexts may be impacted by such changes. 

\section{Interview Findings}
\label{sec:findings}
\begin{figure*}[b]
  \centering
  \includegraphics[width=\linewidth]{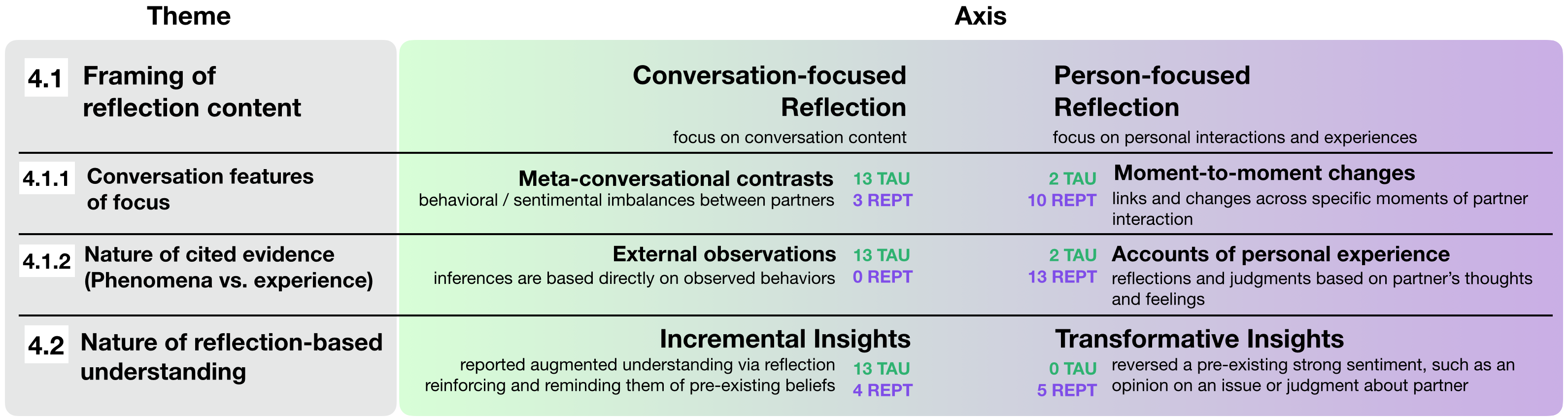}
  \caption{\revisionb{\textbf{Visual guide of each axial theme} from \textit{Partner A}'s reflective interview. In each subsection of Section \ref{sec:findings}, we define criteria for the reflection property in each axial sub-theme and report the counts of participants across REPT and TAU conditions who exhibited each property during the interview; salient differences across conditions are grouped to left and right.}}
  \label{fig:interview}
  \Description{Visual guide of thematic organization and axes for \textit{Partner A}'s reflective interview. We provide salient definitions for each sub-theme and report the count of participants who exhibited each reflection behavior across REPT and TAU conditions.}
\end{figure*}

Our interview uncovered how participants who re-experienced the conflict discussion (\textit{Partner A}) reflected on the conversation and their perceptions of self and partner (\textbf{RQ1}).
We present how qualitative themes \revisionb{(See Fig. \ref{fig:interview} for a visualization)} manifested across participants of the two conditions in two sections. 
First, we identify how participants tended to ground their spoken reflections in either conversational content or personal experiences.
We then describe an observed contrast: multiple participants only from the REPT condition described reversing previously-ingrained beliefs about their partner and themselves, compared to more incremental insights experienced by TAU participants.

\subsection{Conversation-focused vs. Person-focused Reflection}
\label{subsec:granularity}
We identified an theme of conversation-focused vs. person-focused reflection in how participants verbally framed their discussions.
This took the form of two axes:
participants tended to either speak of the conversation as a whole, or center on moment-to-moment changes throughout the conversation with their partner. 
They also tended to fall into the categories of providing either objective or subjective descriptions.
These differences in framing additionally affected distinctions in the types of actionable solutions which they proposed for future conflict discussions.

\subsubsection{Focus on meta-conversation vs. moment-to-moment changes}
\label{subsubsec:meta}
We identified two types of ways in how participants scoped their references to the conversation. Some participants tended to mention conversation-wide trends while reflecting on the discussion they watched, while others tended to cite detailed moment-to-moment changes specific to isolated parts of the conversation. 
We include more detailed descriptions and examples below.

Participants who focused on conversation \textbf{meta-features} often mentioned trends in the form of behavioral or sentimental imbalances that they noticed between themselves and their partner, which directly influenced their goals for future conversations. 
Behavioral imbalances participants noticed included imbalance in power (T1)\footnote{Identifiers of participants from the TAU condition will be prefixed with \textit{T}, and \textit{V} for the REPT condition.}, ratio of talking vs. listening (T3, T6, T13), and level of self-disclosure (T5, V4, V6). 
T5 described his realization of an imbalance in the method of communication: \textit{"I noticed that I'm asking her questions most of the time. And she's speaking most of the time. And I'm not saying a lot without being prompted."} 
C10 described a sentimental imbalance when she spoke of \textit{"the very contrasting emotion [my partner and I] have on this issue,"} as her partner expressed eager interest in the issue while she was feeling tired of explaining herself to him.
Other types of meta-conversational features included realization of positive communication dynamic between partners (T2, T7, T12) and recap of conversation content (T5, T8).

Participants who focused on \textbf{moment-to-moment changes} referenced more links between specific behaviors and the state of the conversation. We define a moment-to-moment change as the explicit recall of changes between specific moments of the conversation, in contrast to general trends across the whole conversation. 
For example, T2 noted \textit{"This is the first time that I expressed worry in the video that I can recall. That was a new emotion and different kind of reaction for me to see."}
Participants also noticed these changes by reflecting on instances of chain reactions between their own and their partner's behavior.
V9 described a lack of change across moments of the conversation: \textit{"I realized this was like the second question back to back that I asked, and I didn't let her ask anything. I felt kind of sad because it seemed like she had a lot of questions too."} 
V13 recalled how seeing those momentary changes helped him understand his partner's frustration: \textit{"Being able to see from her perspective what transpired just before this moment, I could tell what I was reacting to... I can see the cause and reaction of [my partner's] frustration, and how it was tied to my lack of totally honest communication. It was helpful to see how I was expressing, and then link to how she was feeling."} 

Overall, meta-conversation reflections resulted in straightforward goals for future conversation such as adjusting for amount of behavior (e.g., giving partner more space to talk, or self-disclosing more actively).
The more subtle momentary differences were linked to more emotionally-detailed goals, such as attending to partners' in-the-moment feelings, as well as incremental mood changes to help maintain positivity and constructivity in the conversation.
We also found a surprising contrast in numbered differences of participants who mentioned one or more of each reference type across conditions: all 13 TAU, and 3 REPT (VR) users explicitly mentioned meta-conversation trends during their reflection, while 2 TAU and 10 REPT users explicitly mentioned moment-to-moment changes.

\subsubsection{Focus on external observations vs. personal experience}
We also identified that some participants provided mostly objective, external observations while others provided subjective, personal descriptions during reflection. 
Objective descriptions included reflections supported by recall of factual, external events which happened during the conversation, or the content of partners' statements. Subjective descriptions included discussion of partners' feelings, thought processes, and perspective. 


In reflections informed by \textbf{objective, external descriptions}, participants tended to make inferences directly based on observed behavior without much further interpretation.
For example, T12 used the following to support his interpretation that his partner wasn't happy with his behavior: \textit{"She said, 'I didn't feel like you were listening to what I was saying.' She said she had to repeat herself."} 
T3 made a similar statement when explaining an insight on her partner's behavior: \textit{"She was just describing how she felt, about our scope of what is and isn't a big deal in terms of conflict."}
T7  used body language to directly infer her partner's intention, describing her partner \textit{"fidgeting a lot and not looking directly at me. So I can tell he wants to say what he wants but is trying to spare my feelings."}
In these cases, conclusions were made as a direct result of the observation without supporting them with interpretations of their partner's internal state.

In contrast to the above, reflections informed by descriptions focusing on \textbf{subjective personal experiences} involved deep discussion of partners' thoughts and feelings in context of the conversation moment.
Juxtaposed with T7's above statement about her partner, V10 reflected on his observation of his partner fidgeting by talking about her emotional state: \textit{"Because I noticed her fidgeting with her fingers, despite me perceiving how confident she was in asking these questions, I realized there is an underlying level of nervousness or insecurity, and fears."}
We observed this focus on subjective experience taking place often in REPT participants due to seeing things from the partner's perspective. In a reflection where V8 said, \textit{I can see why it would be hurtful towards her to not get up and hug her or ask her what's wrong,"} she cited the reason as, \textit{"I think that seeing [my partner's] perspective showed that it is a hurtful thing I've been doing."} 
V11 also described her partner's feelings and thoughts in succession when reflecting: \textit{"Watching it back, I realized he really did make a choice to continue our relationship. Even though there were a lot of unknowns, he made the choice to stay in my life, even though it probably looks really scary."} When making reflective decisions, these individuals spoke in detail of their partners' subjective experiences instead of focusing on what they did. 

Between individuals who attended to external, objective content vs. subjective personal experiences, we saw that reflections based mainly on straightforward observations resulted in goals of controlling externalized behaviors.
Conversation goals that stemmed from descriptions of subjective experience tended to be more sentiment-focused: these participants chose to embrace honest self-expression and more openness toward their partner.
In numbering coded occurrences of our operationalized objective and subjective description in participants across conditions, we again found a surprising divergence: all 13 TAU and 0 REPT participants made reflections based on objectively-stated observations, while 2 TAU and all 13 REPT participants made reflections based on subjectively-stated observations.


\subsection{Incremental vs. Transformative Insights}
\label{subsec:insights}
In this section, we detail various types of `\textit{transformative insights}' which were experienced exclusively by REPT users in our study. We define transformative insights as changes in opinion or attitude which participants described as completely reversing a strongly-held sentiment they previously had. These contrast with what we refer to as \textit{incremental insights}, ones which participants described to augment, confirm, or remind of their pre-existing understandings. 

\textbf{\textit{Incremental insights}} occurred in the great majority of our TAU participants, who gained augmented understanding through video review of the discussion by reinforcing certain beliefs or reminding them of important conversation events they otherwise would have forgotten.
Many of the participants found it important to note and remember significant points of \textbf{disagreement/agreement} (T2, T6-7, T11-13), the nature of the \textbf{conversation/conflict dynamic} with their partner (T2, T4-5, T7-10, T12, V4, V6) (see \autoref{subsubsec:meta}), explanations of own/partner's \textbf{viewpoints} (T3-4, T8, V4), and demonstrations of \textbf{partner attitude} (T1, T4, V2). 
These insights were cited as useful for maintaining awareness of dynamics to prevent or promote in future conversations.

\textbf{\textit{Transformative insights}} reversed several users' strongly-held opinions relating to their partners, which they validated with thorough explanations of their thought process.
Below, V3 discussed how taking her partner's perspective reversed her feelings about a long-term behavioral habit which had frustrated her throughout their decade-long marriage.
\begin{quote}
"I found a lot of value in watching his hands. My husband does a lot of repetitive hand movements when he's nervous, and it tends to frustrate me, and make me feel like he is uncomfortable with what I'm saying. \textbf{Watching him do it from his perspective, I felt uncomfortable vs. frustrated.} Seeing myself talk to him the way I did, \textbf{I can now understand why he would make those kinds of gestures} because even `I' was nervous with how absolute and sure I was when speaking to him."
\end{quote}
V3 explains how the embodied perspective-taking linked to re-evaluating her original perspective.
Her reflection on how her partner experienced her own behavior tied into another belief change about herself:
\begin{quote}
    "I think my biggest realization is that I thought my husband was the major reason that we had trouble communicating. And while he might not like conflict, I spend a lot of time saying what he's doing, versus what I'm doing. \textbf{I have taken this approach to this conversation so many times}, and hearing/watching myself from this point of view makes me think about how many times my partner has been on the receiving end of me pointing out things and for me, doing that \textbf{it felt like, \textit{here we go again}, but not from my standpoint, from his standpoint — of like, here she goes again.}" 
\end{quote}
In contrast to her previous belief about her husband, V3 developed a self-perception that was the opposite. Her future conversation goals also included giving her partner more opportunities to express himself. 

V1 presents another interesting transformative insight where the participant's opinion on the conflict issue is reversed. In the study, V1 and his wife discussed a financial conflict: he believed in separated finances and her in joint finances. After the REPT experience, he detailed a thought process in which his opinion changed to match his wife's:
\begin{quote}
    "When I was watching the video in her seat, I thought yes, we could share money also, because we shared our household things, and I help her in things I am an expert in and she does the same. So as we are sharing everything, why not money? \textbf{As I sat in her place while she was telling me this, \textit{I thought: I am saying these things to myself.} I considered these thoughts as if they were mine,} and in that particular point, she was right. Here I realized that there is a different way to think about this, and we can also share our money. "
\end{quote}
After embodying her perspective, V1 reconsidered his wife's expressed viewpoints as if they were his own. He said later on, \textit{"I thought she was arguing with me, but while watching this I saw that she was just giving an opinion that is similar to mine so I can think about that perspective."} The melding of perspectives led to this fundamental change in opinion.

Other forms of transformative insights included changes in self-perception triggered by perspective-taking.
Some REPT participants experienced cognitive dissonance when watching themselves say something they realized they did not agree with completely. For example,
V10 stated, \textit{It was a moment where I realized oh, I don't actually want that. I said what I wanted, that's not something I want anymore. Hearing these words that I wouldn't say again made me even more certain of it."}
Others changed an aspect of their self-image based on what they imagined their partner would believe about themselves:
V11 originally believed that she was not committed to the relationship, \textit{"[but] watching it all from his perspective, I almost wonder if that's not exactly true. I think that if I really didn't do enough for him, I don't think he would stand for that. I think he would probably have broken up with me."}

Many user accounts implied that the feeling of embodied perspective-taking heavily contributed to these transformative insights. This tied to feelings of close connection with one's partner, which participants stated were triggered by specific embodiment cues.
V11 described her REPT experience in the following manner:
\begin{quote}
    "This experience leads to so much self reflection. Looking down and realizing like, his hands are right there. There are some times where he turned his head. And I turned my head, while he was looking. And it's almost like I'm controlling the movements, but I'm actually like him. I think it gave me a new sense of empathy, and \textbf{dive deeper by imagining how they feel and what they're thinking, especially when you basically are them, or pretending to be them, and seeing what they see and hearing what they hear.} Having the person on the other side be someone who means so much to you made it feel more personal, and made me think a lot about his perspective."
\end{quote}
The above-mentioned contribution of \textbf{head movement} to the sense of perspective-taking and embodiment of their partner's experience was cited by a great majority of the participants (V1, V4, V6-7, V10), and supported reflections exclusively in the REPT condition. 
V3 delivered another account of this connection:
\begin{quote}
    "It's like a mirror, to be able to see myself, but little things, like my husband's head moving. And not necessarily knowing why his head moved, but just getting that sense of being in his view. \textbf{Being able to see the environment move} was impactful. Sometimes the environment would move when I was really continuing to talk. And that to me, showed that it had an impact on him."
\end{quote}

\section{Quantitative Results}
\label{sec:results}
\revisionb{To address our research goals of understanding how the interventions affect both outcomes of participants' social perceptions and communication behavior, we employed two forms of quantitative measures to enable triangulation of our qualitative findings (RQ1) and to fully address RQ2.}
\revisionb{To evaluate user reflection across \textit{TAU} and \textit{REPT conditions} (\textbf{RQ1}), we complemented our qualitative in-depth analyses of users' subjective reflections on their interactions, selves, and partners (Sec. \ref{sec:findings}) with measurement of affective empathic accuracy \cite{schulz2004looking} as a quantitative element of user reflection.
Our assessment of users' subsequent communication behavior (\textbf{RQ2}) is fully reflected through use of the multi-dimensional IDCS measure of dyadic communication behavior to form comparisons across the two conditions.}

\subsection{Empathic Accuracy}
\label{subsec:empathicaccuracy}
We analyzed the empathic accuracy across conditions by calculating the sum of the five affect ratings by Partner A and B in both conditions, then conducting two-tailed t-tests on the differences.
Results from the t-test on Partner A ($t(24) = 0.76$, $p = 0.46$) and Partner B ($t(24) = 0.068$, $p = 0.95$) did not indicate significant differences at $p\leq 0.05$.

\begin{table}[htb!]
\caption{Empathic Accuracy (Mean/SD of Summed Diffs)}
\label{tab:empathicaccuracy}
\centering
\setlength{\tabcolsep}{10pt} 
\renewcommand{\arraystretch}{1.2} 
\small
  \begin{tabular}{lll}
    \toprule
    \textbf{Condition} & \textbf{Partner}&\textbf{Empathic Accuracy (Sum)}\\
    \midrule
    REPT & \textsc{Partner A} & $-4.0$ ($6.32$)\\
    TAU & & $-1.85$ ($8.09$)\\
    \midrule
    REPT & \textsc{Partner B}& $-4.0$ ($4.16$)\\
     TAU & & $-3.85$ ($6.95$)\\
  \bottomrule
  \end{tabular}
\end{table}
\subsection{Interactional Dimensions (IDCS)}
\label{subsec:IDCS}
\begin{figure*}[htb!]
     \centering
     \begin{subfigure}[b]{0.495\textwidth}
         \centering
         \includegraphics[width=\textwidth]{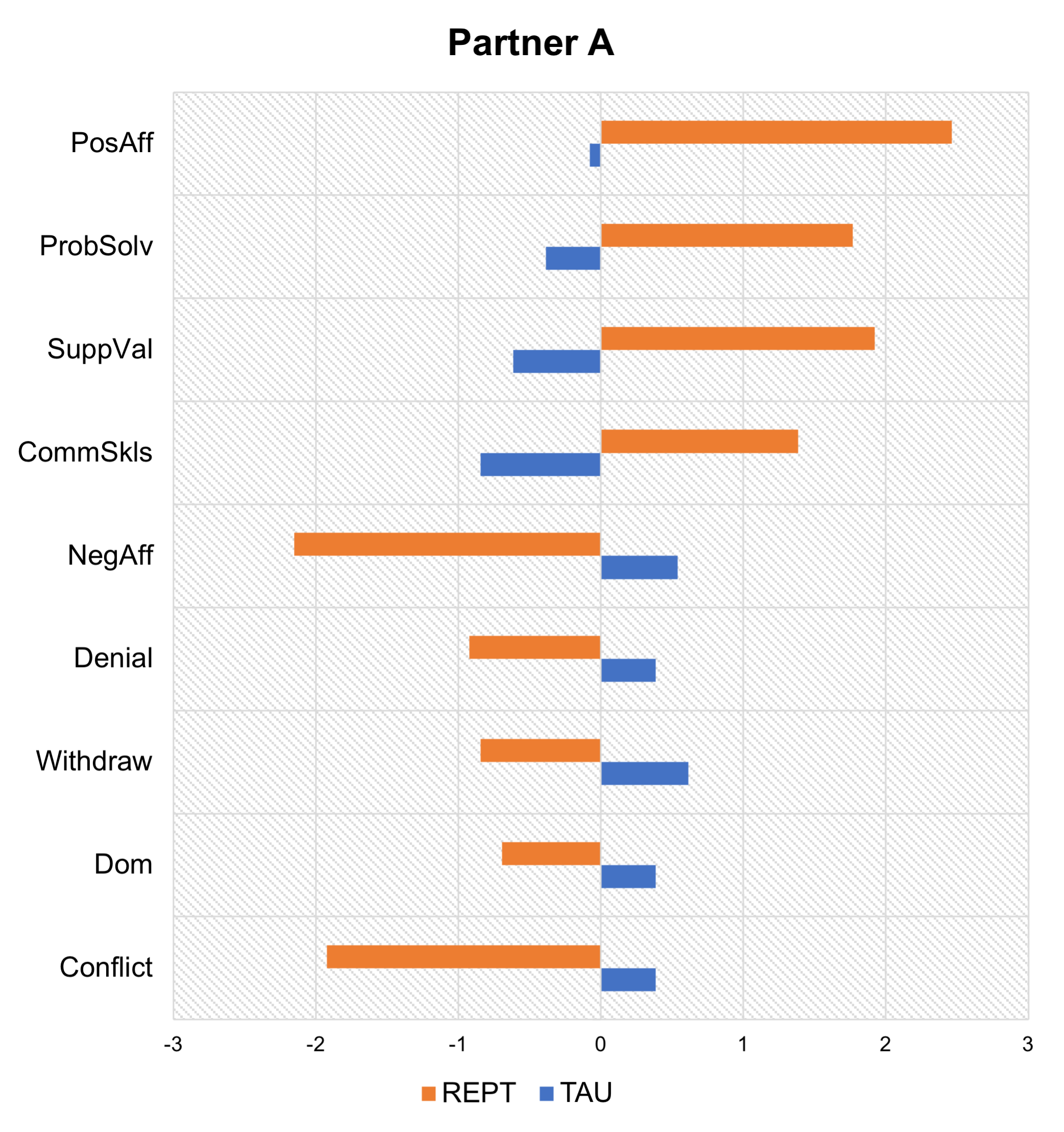}
     \end{subfigure}
     \begin{subfigure}[b]{0.495\textwidth}
         \centering
         \includegraphics[width=\textwidth]{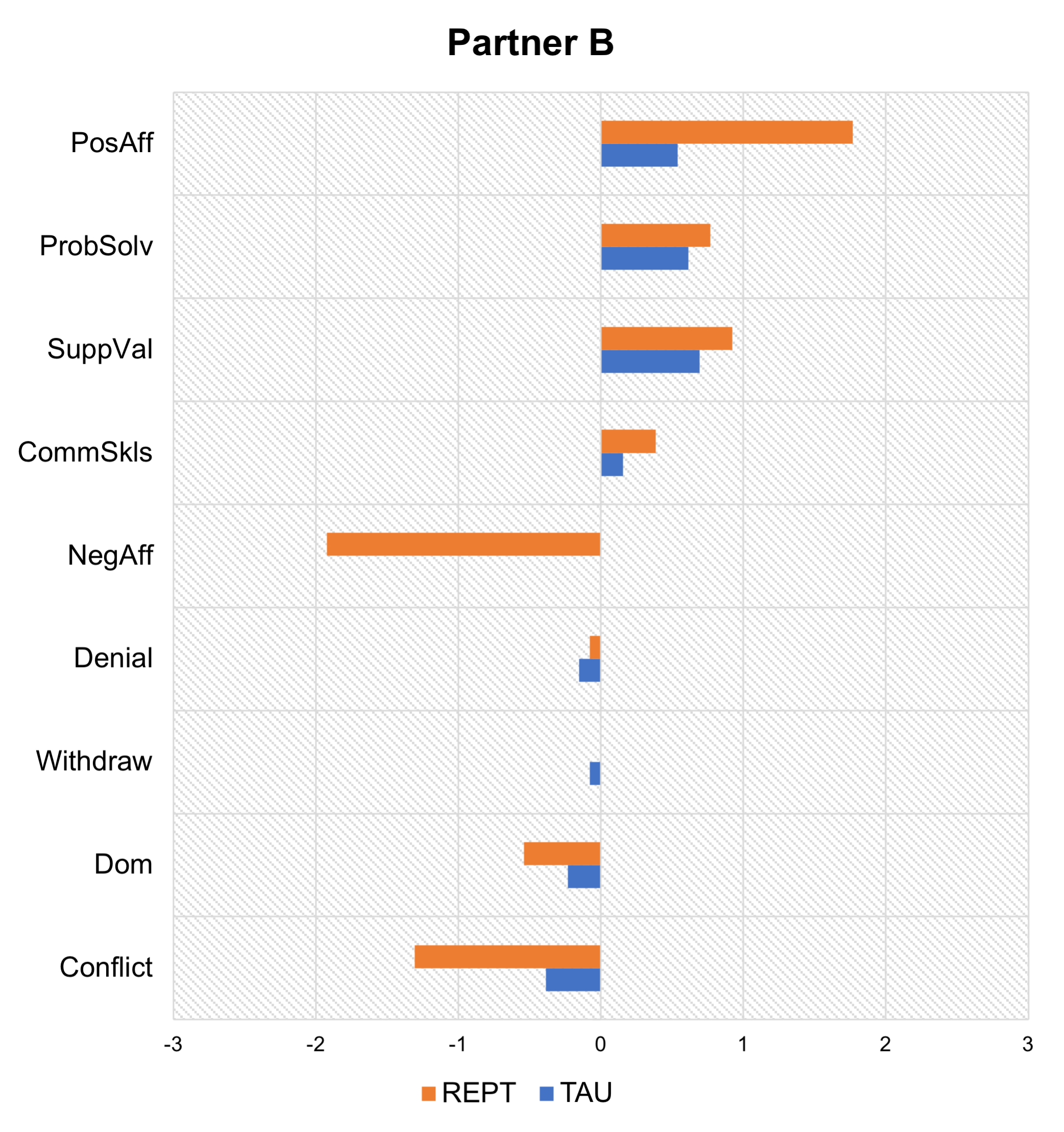}
     \end{subfigure}
     \caption{\revisionb{\textbf{Mean change in IDCS individual dimension scores from Discussion 1 to Discussion 2}, shown across REPT and TAU conditions separately for \textit{Partner A} and \textit{Partner B} (See \autoref{fig:procedure} for study procedure). X-axis specifies the change in score; Y-axis specifies IDCS individual dimensions in the following top-down order: Positive Affect, Problem Solving, Support/Validation, Communication Skills, Negative Affect, Denial, Withdrawal, Dominance, Conflict (See \autoref{tab:idcs} for definitions).}}
     \Description{\revisionb{\textbf{Mean change in IDCS individual dimension scores from Discussion 1 to Discussion 2}, shown across REPT and TAU conditions separately for \textit{Partner A} and \textit{Partner B} (See \autoref{fig:procedure} for study procedure). X-axis specifies the change in score; Y-axis specifies IDCS individual dimensions in the following top-down order: Positive Affect, Problem Solving, Support/Validation, Communication Skills, Negative Affect, Denial, Withdrawal, Dominance, Conflict (See \autoref{tab:idcs} for definitions).}}
     \label{fig:IDCSresultschart}
\end{figure*}

Prior to assessing our main measures, we conducted two-tailed t-tests to compare the starting IDCS scores (from Discussion 1 in Session 1) across the two conditions in order to determine if the REPT technology setup impacted communication behavior. No significant differences were found for either partner in the above tests. We then proceeded to assess the main measures.
\revisionb{For each dimension in IDCS (see \autoref{tab:idcs}), we conducted two-tailed t-tests to perform a cross-condition comparison of the change in score for each of the 11 dimensions between the pre/post conflict discussions.} Positive values indicate an increase in the dimension score between the first and second conflict discussion, while negative values indicate a decrease in the dimension score.
Results from the t-tests indicated significant differences between conditions for both \textit{dyadic dimensions} (at $p\leq 0.01$): \textbf{positive escalation} ($t(24) = -3.13$, $p = 0.005$, \revisionb{$d=1.23$}) and \textbf{negative escalation} ($t(24) = 3.93$, $p<0.001$, \revisionb{$d=1.54$}). Calculation of Cohen's $d$ found a medium effect size. Below are means of the dyadic measures across conditions; In the REPT condition, positive escalation significantly increased, while negative escalation significantly decreased (\autoref{tab:idcsdyadic}).
\begin{table}[htb!]
\caption{IDCS Dyadic Dimensions - Mean/SD of Differences}
\label{tab:idcsdyadic}
\centering
\setlength{\tabcolsep}{10pt} 
\renewcommand{\arraystretch}{1.2} 
\small
\begin{tabular}{lll}
    \toprule
    \textbf{IDCS Dyadic Dimension}&\textbf{REPT}&\textbf{TAU}\\
    \midrule
    Positive Escalation&2.69 (1.75)&0.15 (2.34)\\
    Negative Escalation&-1.62 (1.56)&0.69 (1.44)\\
  \bottomrule
\end{tabular}
\end{table}
\\
T-tests were conducted separately for \textit{Partner A} (who experienced the retrospection intervention) and \textit{Partner B} to measure changes across conditions in the nine \textit{individual dimensions}. For \textit{Partner A}, results from the \revisionb{t-tests} indicated significant differences (at $p\leq 0.05$) for all except two dimensions: \textbf{conflict} ($t(24) = 3.145$, $p = 0.004$, \revisionb{$d=1.23$}), \textbf{denial} ($t(24) = 2.65$, $p = 0.014$, \revisionb{$d=1.04$}), \textbf{communication skills} ($t(24) = -3.22$, $p = 0.004$, \revisionb{$d=1.26$}), \textbf{support/validation} ($t(24) = -4.58$, $p<0.001$, \revisionb{$d=1.79$}), \textbf{problem solving} ($t(24) = -3.06$, $p = 0.005$, \revisionb{$d=1.20$}), \textbf{negative affect} ($t(24) = 4.173$, $p<0.001$, \revisionb{$d=1.64$}), and \textbf{positive affect} ($t(24) = -3.96$, $p<0.001$, \revisionb{$d=1.55$}).
Calculation of Cohen's $d$ found a medium effect size across all of the above.
From \autoref{fig:IDCSresultschart}, it is shown that conflict, denial, and negative affect significantly decreased, while communication skills, support/validation, problem solving, and positive affect increased in \textit{Partner A} over the course of the study.
Significant differences were not found for the dominance and withdrawal dimensions in \textit{Partner A} scores.
For \textit{Partner B}, significant differences were found only for \textbf{negative affect} ($t(24) = 2.39$, $p = 0.025$, \revisionb{$d=0.94$}). The mean value for \textbf{positive affect} change in \textit{Partner B} was considerably high (\autoref{fig:IDCSresultschart}), but the t-test results were short of significant: ($t(24) = -1.79$, $p = 0.086$, \revisionb{$d=0.70$}). Mean IDCS score changes are shown in \autoref{tab:idcsindividual}.

\revision{Our IDCS results answer \textbf{RQ2} by showing how multiple dimensions of subsequent conflict communication improved significantly in REPT over TAU. In \textit{Partner A}, we saw an increase in positive qualities such as communication skills, support and validation, and problem solving, accompanied by decrease in negative dimensions of conflict and denial (\autoref{fig:IDCSresultschart})}. Both \textit{Partner A} and \textit{Partner B} in REPT reflected higher positive affect and lower negative affect in relation to TAU, which also corresponded to the changes in dyadic positive and negative escalation. Based on this pattern of difference across partners in the REPT condition, it is possible that the changes in \textit{Partner B}'s affect related to positive changes seen in \textit{Partner A}'s communication behavior.

\section{Discussion}
\label{sec:discussion}
\revisionb{The following sections will cover three areas: we summarize the principal findings of how our RQs were addressed, discuss how our findings on REPT extend the literature and present new ways to bridge contrasting human perspectives, then propose two design implications that form an agenda for how representations of felt experience can be integrated into technologies for social reflection: \textit{embodied social cognition} and \textit{embodied experience as interaction context}.}

\begin{table*}[htb!]
\caption{IDCS Individual Dimensions - Mean/SD of \revisionb{Difference Across Sessions}}
\label{tab:idcsindividual}
\centering
\small
\setlength{\tabcolsep}{10pt} 
\renewcommand{\arraystretch}{1.2} 
\begin{tabular}{lrrrr}
    \toprule
    \textbf{IDCS Individual Dimension}&\textbf{REPT}&\textbf{TAU}&\textbf{REPT}&\textbf{TAU}\\
    \midrule
     & \textsc{Partner A}& &\textsc{Partner B} & \\
    Conflict & -1.92 (1.98) & 0.38 (1.76) & -1.31 (1.25) & -0.38 (2.29)\\
    Dominance & -0.69 (1.93) & 0.38 (1.61) & -0.54 (0.97) & -0.23 (1.09)\\
    Withdrawal & -0.85 (2.23) & 0.62 (2.47) & 0.00 (1.00) & -0.08 (2.59)\\
    Denial & -0.92 (1.55) & 0.38 (0.87) & -0.08 (0.28) & -0.15 (0.55)\\
    Communication Skills & 1.38 (1.33) & -0.85 (2.12) & 0.38 (1.19) & 0.15 (1.21) \\
    Support/Validation & 1.92 (1.19) & -0.62 (1.61) & 0.92 (1.11) & 0.69 (1.11) \\
    Problem Solving & 1.77 (1.36) & -0.38 (2.14) & 0.77 (1.74) & 0.62 (2.43) \\
    Negative Affect & -2.15 (1.63) & 0.54 (1.66) & -1.92 (1.12) & 0.00 (2.68)\\
    Positive Affect & 2.46 (1.61) & -0.08 (1.66) & 1.77 (1.01) & 0.54 (2.26) \\
  \bottomrule
  \end{tabular}
\end{table*}

\subsection{Principal Findings}
\revisionb{The current work presents a novel \textit{meta-conflict} approach to extend the scope of conflict resolution technology (see \ref{subsubsec:conflict}) by using VR other-embodiment to facilitate reflection on conflicting perspective. Novel in the study of VR embodied cognition and empathy (\ref{subsubsec:embodiment}), we also demonstrate previously-unidentified social effects in the context of ``\textit{familiar others}.''
In response to \textbf{RQ1} which asked how REPT (vs. TAU) impacted users' reflections on the discussion and perceptions of themselves and their partner, we showed (Sec. \ref{sec:findings}) that users who embodied their partners in REPT reflected more on momentary, subjective experiences; developing changes in perception of the conflict issue itself and of their relationship with their partner (e.g., communication style). TAU users reflected more on bigger-picture discussion dynamic and did not report changed perceptions.
\textbf{RQ2}'s inquiry of how REPT (vs. TAU) affects conflict communication is answered through our pre/post change analysis of IDCS (\autoref{fig:IDCSresultschart}) showing that REPT results in significantly higher improvements on interactional dimensions such as communication skills, problem solving, support and validation, positive affect, and significantly higher decreases in negative affect, conflict, and denial.}

\subsection{Across Methods: Resolution of Perspectives Through Embodied Understanding}
\textit{Intersubjectivity} (sharing understanding) is a theme that runs throughout this work.
Combined insights from the qualitative interview findings and quantitative communication measures demonstrate that \textit{Partner A} in the REPT condition, who experienced embodied perspective-taking of their partner, was able to develop major insights (Sec. \ref{subsec:insights}) about themselves and their partner and experience significant improvements in conflict interaction (Sec. \ref{subsec:IDCS}).

Retrospective embodied perspective-taking (REPT) enabled close others to immerse in their partner's lived experiences and explore competing personal perspectives.
In-the-moment details (Sec. \ref{subsec:granularity}) reflected realizations about interactions, which led to more candid self-expression and appreciation for shared experience with their partners.
Transformative insights (Sec. \ref{subsec:insights}) illustrated major re-evaluation of existing perspectives, which is particularly significant in light of the work on various social biases in close others \cite{kenny_accuracy_2001,nickerson_how_1999,thompson_judgments_1981}, and the failure of traditional media reflection methods (TAU) to mitigate these biases \cite{fichten_videotape_1983}.
\textbf{In a sense, the REPT method can be a facilitator of intersubjectivity by enabling perspectives of both the self and other to be embodied in tandem.}
Our interview findings showed individuals re-evaluating their partner's competing opinion on an issue as if it was their own, to individuals reassessing their own self-image after seeing themselves through their partner's eyes.

Our IDCS measures demonstrated evidence of REPT as a reflection intervention which improved communication between close others, supplementing TAU \cite{fichten_see_1984,fichten_problem-solving_1983,fichten_videotape_1983} with an alternative approach. 
\revisionb{Our findings also contribute formative evidence of a link between virtual embodiment and \textit{content-grounded} \textbf{cognitive empathy}, which is a previously-unobserved effect in the field of VR-embodied cognition.} A 2021 meta-analysis \cite{martingano_virtual_2021} shows limited research linking VR embodiment to cognitive empathy, which requires more effort and is harder to stimulate than emotional empathy, but provides the crucial pathway to meaningful changes in social behavior \cite{jiang_differing_2021,georgiou_what_2019,gambin_relations_2018}.
Though lacking measureability, our interview content appears to contain examples of cognitive empathy (content detailing specific other's thoughts and experiences) evoked by embodied perspective-taking, at a level which was strong enough to stimulate transformative insights in users.
Such formative evidence suggests that embodied experiences could be applied to stimulate cognitive empathy as a pathway for promoting meaningful and intent-driven prosocial behavior amidst contrasting perspectives, which is relevant in many social contexts (e.g., cross-cultural collaboration \cite{bochner2013cultures}, diversity and inclusion \cite{ferdman2017}, public policy \cite{greener_more_2010}).
Further exploration can lead to more nuanced insights on designing embodied experiences that can induce cognitive empathy and perspective-resolution in different contexts.

\subsection{Embodied Social Cognition: Integrating Felt Experience With Creative Reasoning}
\textit{Embodied social cognition} is also linked to our findings through virtually-embodied experience and the mental processes of perceiving and engaging with another person.
Originating from the cognitive science domain \cite{gallagher_22_2008}, it shares roots with the HCI principles of \textit{embodied cognition} \cite{kirsh_embodied_2013} and \textit{embodied interaction} \cite{dourish_where_2001} that advocate against disembodiment of rationality.
In the same spirit, we propose that embodied social cognition can be extended from these principles as a new approach towards designing socially reflective technology which prioritizes embodied social rationality. 
Embodying human experiences can help people creatively \textit{generate} useful social understandings, which is important given the distinctness of social context; it is different from more tangible or objective contexts in that there is no true way to `accurately' access the inner state of a person. 
\textbf{Our findings are an example of how the embodiment process can support \textit{creative social cognition} and \textit{self-generated} understanding by stimulating original thoughts about the self and other which are independent of any information `directly' presented to the user.}

So far, existing HCI and social informatics work on interpersonal social cognition (perceiving and understanding the thoughts and experiences of specific people) has mostly emphasized \textit{information-centered approaches}, focusing on how external representations of social information can be best observed and understood (e.g., narratives, biodata, explicit social cues, communication media) \cite{cosley_artlinks_2008,griggio_augmenting_2019,significantotter,birmingham_can_2020,miller_understanding_2017}.
In contrast, embodied social cognition emphasizes how an \textit{experience-centered approach} instead of observed knowledge can contribute to effective, in-depth sensemaking and reasoning about other people. The role of mirror neurons in embodied social simulation (exposure to perceptual stimuli triggering brain activations for corresponding social actions) provides a unique psycho-physical link between these experiences and their ability to stimulate vivid thoughts about other people specifically \cite{spaulding_introduction_2012,gallese_what_2011}.
This shows that experiential channels, and not just information-centered ones, are powerful enough to support the creation of cognitive social understanding and not just basic emotional experiences such as affective empathy \cite{herrera_building_2018,martingano_virtual_2021}.

Technology-embodied experiences can then be applicable to use cases more complex than previously imagined (singular empathy-inducing sessions), such as integrating a reflective step into a pipeline of social decision-making, or support for learning context-specific social behaviors.
Virtually-embodied social cognition also has potential to reconcile a known duality between information-rich and information-poor approaches for social understanding. Historically, observing social cues (information-rich) is known to conflict with the ability to reason about another's subjective experience, but raw attempts to take someone's perspective (information-poor) projects biases instead of improving understanding \cite{vorauer_potential_2013}.
Embodied social cognition could resolve this by turning the originally information-poor perspective-taking process into an immersive, experience-rich process which replaces originally-observed cues by ones which can be felt and experienced, providing the scaffolding necessary to creatively generate social understanding. 
Experience is as good as knowledge: future work can integrate more felt- or embodied- elements (e.g., VR immersion, or biosignals \cite{liu_effect_2019,costa_regulating_2018}) into social reflection tools, which could better facilitate self-constructed understanding of a social perspective or experience instead of attempting to digestibly present an array of information for the user to observe.

\subsection{Advancing Social Reflection: Embodied Experience as an Interaction Context}
Across the TAU and REPT conditions we saw a contrast in attention to more high-level vs. experience-specific qualities of the social interaction. 
This shows that the medium of interaction ``context" (in the current study, desktop video or immersive VR) can frame the social context of how the user thinks and reflects upon their experience.
In the VR embodiment implementation, we also showed that users were able to manually mark important parts of the interaction while virtually embodying their partner (\autoref{subsec:procedure}) without breaking the level of immersion required to generate notable social insights (\autoref{subsec:insights}). These manually-selected moments are heavily linked to the reflections reported in our findings, which were largely elaborated explanations of those selected moments.
\textbf{Given that such ``meta-experiential" interactions can be integrated into VR perspective-taking, social virtual embodiment can be not just an isolated experience (its current state of application) but also an \textit{interaction context}: an environment which we can build upon to further advance user engagement in social reflection.}

The potential for more advanced levels of interfacing with virtually-embodied perspective-taking can allow people to engage in more complex forms of analysis or decision-making while still in the valuable altered frame of mind that immersive social embodiment provides. 
In current approaches in VR perspective-taking for empathy, users are completely inundated during the embodied experience and only have space to digest afterwards, at which point their mental context may already be different.
Being able to analyze a social interaction while still in the embodied state may be useful in more complex social contexts that necessitate understanding interactions between multiple human perspectives/experiences, or making sense of different elements in the conversation or issue at hand. 
For example, someone may try to understand interacting perspectives in a three-way conversation or reflect on a discussion with multiple disparate elements.
Potential ``meta-experiential" interactions can support switching between socially-embodied perspectives, enable passive annotation, or allow reviewing selections of an embodied interaction at one's own pace. 

One potential solution to the difficulty of balancing potentially-disruptive interactions with the immersive social embodiment in another's perspective is to design natural transitions that are consistent with people's semantic and sensory understanding of conscious experience. Existing work in VR has demonstrated the importance of scene transitions and metaphors for maintaining users' experience, awareness, and sense of place \cite{rahimi2018scene,husung2019portals,wang2020slice}; natural transitions can stay consistent with conscious experience by simulating corresponding bodily states such as dreaming (e.g., prompting the user to close their eyes and awaken in another ``state of consciousness,").
Exploring ways to build upon embodied experience as an interaction context can be particularly valuable when multiple people have fundamentally disparate understandings that are rooted in lived experience (e.g. cultural differences, social roles), potentially necessitating analysis of various perspectives and elements of the issue.

\section{Limitations and Future Work}
The need to do exploratory work for a new area means that our mixed-methods study identified more formative insights such as the basic impacts of REPT on users instead of providing targeted statistical evidence.
Future work building on the potential applications and directions for studying this technology can employ more precise inquiry methods to confirm specific hypotheses that arise from this work.
For example, our affective empathic accuracy measure (\autoref{subsec:empathicaccuracy}) could not provide insights on the accuracy of participants' perceptions.
However, past work has shown that 'underestimating' a close other's thoughts actually improves constructive communication \cite{vorauer_potential_2013} and future investigation may want to clarify the nuanced role of reflection accuracy in mediating the behavioral impacts of REPT.
\revisionb{Though our power analysis (\ref{subsec:power}) showed low likelihood of any empathic accuracy effect, we provide future sample sizes to detect or replicate other effects from this study. A sample of $N=16$ is well-powered to replicate our main IDCS effects, and $N=44$ may detect medium effects for the two IDCS dimensions where we did not find significance (dominance and withdrawal).}
The transformative insights in this work were also one of the most interesting findings - we may want to identify what individual or experiential factors lead to this phenomenon, or characterize a fuller scope of what attributes constitute a transformative insight.
Longitudinal investigations and ecologically-valid settings (e.g., a therapy context) may also be important to determine the longer-term and in-the-field impacts of REPT. 

\revisionb{Our formative findings provide limited nuance regarding personalization and key perceptual mechanisms of the REPT system, necessitating more targeted lines of investigation.}
\revisionb{We explored a socially-complex and highly subjective phenomenon: perception and communication between intimate others who have conflicting perspectives. Individual variability in factors such as personality, attachment, communication style, and conflict context can have complex impacts on interaction between highly-interdependent people. Although this initial work did not find any significant impact from the individual traits we measured, refining REPT's intended effects for a broader audience requires more targeted sampling and nuanced assessment of how social and individual differences affect user responses.}
\revisionb{Our work also prioritized applicability, and we created REPT to be a consumer-accessible and reproducible system. However, more detailed exploration of perceptual affordances such as those across our conditions (e.g., immersivity, user agency, perspective breadth) may provide insights on how couples’ responses relate to specific aspects of this technology and inform more effective integration into future systems. These response effects may be motivated by distinct sociocognitive theories (embodied cognition vs. facial attention) and can be explored across multiple lines of work.}

\revisionb{We also consider potential lines of future research which stem from our design discussions and the novel contexts explored in our work.}
Embodied social cognition through technology can explore flexible multisensory methods (e.g., combinations or subsets of haptic \cite{ju_haptic_2021}, sonic \cite{winters_can_2021}, olfactory \cite{ranasinghe_season_2018}, and visual \cite{jarvela_augmented_2021} technologies) if future research sheds more light on how specific perceptual cues from virtual embodiment relate to social cognition \cite{kilteni_sense_2012,han_immersive_2022,calabro_understanding_2020}. We saw hints of this from REPT participants referring to specifically head and hand movements as triggering cues in our findings. 
Considering embodied experience as an interaction context can also lead to development of consciousness-oriented transition methods for VR (to complement the current spatially-oriented ones) as well as novel interaction methods which build on embodied perspective-taking.
\revisionb{To expand VR work on ``\textit{intimate social contexts},'' investigating how patterns of responsivity to VR embodiment might vary across different levels of \textit{\textbf{relational proximity}} (self/close others/strangers) can contribute a new dimension of theoretical knowledge to embodied cognition in VR.}
\revisionb{Building on our demonstration that \textit{meta-conflict} reflection still has positive impacts on actual behavior during conflict, future HCI conflict resolution approaches can move beyond embedding interventions inside conflicts (\ref{subsubsec:conflict}) and also consider the overarching effects of applying interventions (e.g., raising awareness of user actions) in-between conflicts.}

\section{Conclusion}
Close relationships are important sources of social support, but prone to high-risk conflict. 
Integrating converging evidence from the domains of HCI, virtual reality, and counseling which demonstrate the potential of virtual embodiment to facilitate constructive perspective-taking in close others, we developed a VR-based retrospective embodied perspective-taking system (REPT) and conducted a mixed-methods evaluation comparing its ability to improve communication during conflict against the current form of traditional video-based reflection.
Our findings demonstrated how REPT resulted in more reflection on lived experiences, fundamental changes in outlook, and improved communication skills during conflict.
Evidence from our results extend past work by demonstrating REPT to correct strong attributional biases in close others, as well as promote cognitive empathy which was previously found to be unaffected by VR experiences.
We also contribute approaches to HCI design. We show through results of the REPT experience that technology-\textit{embodied social cognition} is a viable process to generate meaningful understandings of other people, which is nontrivial due to the intangible nature of others' mental states.
The unintrusive interactivity of the REPT experience also demonstrated the potential to view \textit{embodied experience as an interaction context}, which can lead to developments of novel interaction techniques that build upon embodiment as a social reflection medium.
We hope that these insights can benefit the ability of future systems to facilitate richer social interactions.

\begin{acks}
We would like to acknowledge several individuals whose support was vital to the completion of this work. Our gratitude extends to Jeff Simpson, who provided valuable consultation and relationship research expertise toward the design of this study. We are also very thankful to Georgie Qiao Jin for her provision of transcription resources and involvement in transcribing the interviews for this study. Additionally, we would like to express our appreciation for the contributions of Samuel Adeniyi and Courtney Hutton Pospick, who provided indispensable assistance in conducting our multiple in-person studies. We are very grateful to all those who contributed to this project in various capacities, as your support was essential to help us reach this successful completion.
\end{acks}

\bibliographystyle{ACM-Reference-Format}
\bibliography{sample-base}

\appendix
\section{Supplementary Analyses}
\balance
\subsection{Power Analysis}
\label{subsec:power}

\begin{figure}[h]
  \centering
  \includegraphics[width=0.47\linewidth]{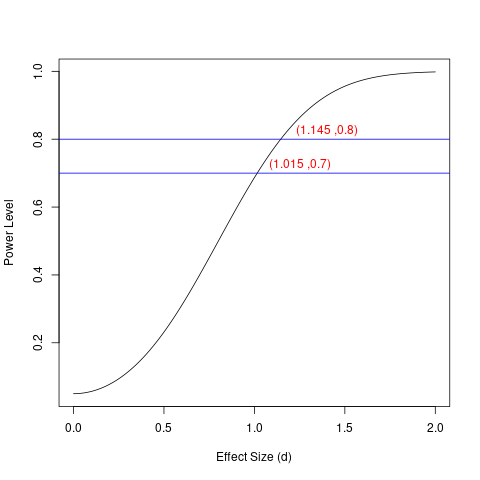}
  \caption{\revisionb{\textbf{Sensitivity power analysis} with our sample size $N = 13$ and significance level $\alpha = 0.05$. With desired power of $0.8$, the predicted effect size under these assumptions is $d = 1.145$. With power $0.7$ the predicted effect size is $d = 1.015$.}}
  \label{fig:sensitivity}
  \Description{\textbf{Sensitivity power analysis} with our sample size $N = 13$ and significance level $\alpha = 0.05$. With desired power of $0.8$, the predicted effect size under these assumptions is $d = 1.145$. With power $0.7$ the predicted effect size is $d = 1.015$.}
\end{figure}

The sample size selection in the current study reflects standard practices in HCI \cite{caine_local_2016}; a priori power analysis was not performed prior to conducting the study due to lack of similar work (VR embodiment effects on specifically empathic accuracy and dimensional conflict behavior) that would allow us to approximate expected effect sizes \cite{sullivan_using_2012}.
Now guided by the effect sizes found in this initial study, we can present a power analysis to inform sample sizes (\autoref{tab:samplepower}) for future work and replication of the current study \cite{sullivan_using_2012}. We also include a sensitivity power analysis (\autoref{fig:sensitivity}) using our current sample size to contextualize the effect sizes and power of our current findings. We do not present a post-hoc power analysis due to its conceptually flawed nature: \cite{dziak_interpretation_2020,yatani_effect_2016,zhang_post_2019}.

The effect sizes found for our significant results on \textit{Partner A} (the main user of interest) were quite large ($>1.0$) (\autoref{tab:samplepower}) and approximated the effect size calculated from the sensitivity analysis at power $0.8$ (\autoref{fig:sensitivity}), showing that the effects we found were quite robust and the current study was relatively well-powered to detect these effects.
However, we provide additional power analyses (\autoref{tab:samplepower}) to help inform sample sizes with sufficient power to either replicate the large effects found in our significant results or potentially detect other effects which we did not find significant in the current work. We follow with recommendations consulting the power analysis in \autoref{tab:samplepower}:

\begin{table}[h]
\centering
\begin{tabular}{l|ll|ll}
\toprule
                 \textbf{IDCS Individual}  & \textbf{Partner A}       &             & \textbf{Partner B}       &             \\
\textbf{Dimension}          & ES ($d$) & $n$ & ES ($d$) & $n$ \\
\midrule
Conflict           & 1.23*           & 12          & 0.500           & 63          \\
Dominance          & 0.61            & 44          & 0.298           & 178         \\
Withdrawal         & 0.62            & 42          & 0.0391          & 10273       \\
Denial             & 1.04*           & 16          & 0.175           & 512         \\
Communication      & 1.26*           & 11          & 0.192           & 428         \\
Support/Validation & 1.79*           & 7           & 0.207           & 366         \\
Problem Solving    & 1.20*           & 12          & 0.0727          & 2969        \\
Negative Affect    & 1.63*           & 7           & 0.938*          & 19          \\
Positive Affect    & 1.55*           & 8           & 0.703           & 33 \\        
\bottomrule
\end{tabular}%

\vspace{1em}

\begin{tabular}{l|ll}
        \toprule
        \textbf{IDCS Dyadic}  & Effect size ($d$) & Sample Size ($n$)\\
        \textbf{Dimension} & & \\
        \midrule
        Positive Escalation & 1.29* & 12 \\
        Negative Escalation & 1.54* & 8 \\
        \bottomrule
    \end{tabular}

    \vspace{1em}

    \begin{tabular}{l|ll}
        \toprule
        \textbf{Empathic}  & Effect size ($d$) & Sample Size ($n$)\\
        \textbf{Accuracy} & & \\
        \midrule
        Partner A & 0.296 & 180 \\
        Partner B & 0.0262 & 22887 \\
        \bottomrule
    \end{tabular}
    \vspace{1em}

\caption{\revisionb{\textbf{Power Analysis - IDCS Dimension \& Empathic Accuracy.} Displayed for each measure are recommended sample sizes to achieve power of 0.8 and reach the effect sizes find in our study, assuming significance level $\alpha = 0.05$. Effect sizes where we found significance in our t-tests are marked with an asterisk*}}
\label{tab:samplepower}
\vspace{-12pt}
\end{table}

Considering that \textit{Partner A} was the main user of interest in our study (as the recipient of the intervention), future studies that wish to replicate the effects we reported for the main user can achieve power 0.8 with a sample size of $N=16$.
Additionally, for \textit{Partner A} we did not observe significant effects for two of the IDCS dimensions: dominance and withdrawal. These still have an at-least medium effect size ($>0.6$) compared to the other dimensions, indicating that a higher-powered study with sample size $N=44$ could potentially detect significant effects for these two dimensions.
Regarding \textit{Partner B}, the current study came close to a significant effect for positive affect ($p=0.086$) with medium-large effect size ($d=0.7$), meaning that a study powered with sample size of $N=33$ or greater may be better-equipped to investigate the secondary effects of REPT on a partner.

\revisionb{Many of the effect sizes for empathic accuracy and \textit{Partner B}'s IDCS measures are quite small, requiring over a hundred participants to reach 80\% possibility of detecting a very small effect. This indicates low likelihood of detecting these effects with the current study design and we do not make sample size recommendations for these variables.}







\end{document}